\documentclass[pra,reprint,a4paper,showpacs]{revtex4-1}
\usepackage{bm}
\usepackage{ulem}
\usepackage[dvipdfmx]{graphicx}
\usepackage{dcolumn}
\usepackage{color}
\usepackage{amsmath}
\usepackage{amssymb}
\usepackage{amsfonts}

\draft % marks overfull lines with a black rule on the right

\begin{document}

%\preprint{}

\title{Time-dependent complete-active-space
self-consistent-field method for atoms: Application to high-harmonic generation
%:High harmonic generation of helium
}



\author{Takeshi Sato}
\email[Electronic mail:]{sato@atto.t.u-tokyo.ac.jp}
\author{Kenichi L. Ishikawa}
\email[Electronic mail:]{ishiken@atto.t.u-tokyo.ac.jp}
\affiliation{
Photon Science Center, School of Engineering, 
The University of Tokyo, 7-3-1 Hongo, Bunkyo-ku, Tokyo 113-8656, Japan
}
\affiliation{
Department of Nuclear Engineering and Management, School of Engineering,
The University of Tokyo, 7-3-1 Hongo, Bunkyo-ku, Tokyo 113-8656, Japan
}

\author{Iva B\v{r}ezinov\'a}
%\email[Electronic mail:]{iva.brezinova@tuwien.ac.at}
\author{Fabian Lackner}
%\email[Electronic mail:]{fabian.lackner@tuwien.ac.at}
\author{Stefan Nagele}
%\email[Electronic mail:]{}
\author{Joachim Burgd\"orfer}
%\email[Electronic mail:]{}
\affiliation{
Institute for Theoretical Physics, Vienna University of Technology,
Wiedner Hauptstra{\ss}e 8-10/136, 1040 Vienna, Austria, EU
}


%\date{\today}

\begin{abstract}
We present the numerical implementation of the time-dependent complete-active-space
self-consistent-field (TD-CASSCF) method 
[Phys.~Rev.~A, {\bf 88}, 023402 (2013)]
for atoms driven by a strong linearly polarized laser pulse. The present
implementation treats the problem in its full dimensionality and
{\color{black}introduces} a gauge-invariant frozen-core approximation, an efficient
evaluation of the Coulomb mean field scaling linearly with the number
of basis functions, and a split-operator method specifically designed
for stable propagation of stiff spatial derivative operators. We apply
this method to high-harmonic generation in helium, beryllium{\color{black},} and neon and explore the role of electron correlations.
\end{abstract}

%\pacs{32.80.Rm, 31.15.A-, 42.65.Ky}% insert suggested PACS numbers in braces on next line

\maketitle %\maketitle must follow title, authors, abstract and \pacs


\section{introduction\label{sec:introduction}}
The rapid progress in experimental techniques for ultrashort optical {\color{black}light} sources with high-intensity has opened new research areas
including ultrafast molecular probing
\cite{Itatani2004Nature,Haessler2010NPhys,Salieres2012RPP}, 
attosecond science
\cite{Agostini2004RPP,Krausz2009RMP,Gallmann2013ARPC}, and
XUV nonlinear optics \cite{Sekikawa2004Nature,Nabekawa2005PRL},
with the ultimate goal to directly measure, and even control
electron motion in atoms, molecules, and solids.
The time-dependent Schr\"odinger equation (TDSE) provides the rigorous
theoretical framework for investigating electron dynamics
\cite{Pindzola1998PRA,Pindzola1998JPB,Colgan2001JPB,
Parker2001,Laulan2003PRA,Piraux2003EPJD,Laulan2004PRA,ATDI2005,
Feist2009PRL,Pazourek2011PRA,He_TPI2012PRL,Suren2012PRA,He_TPI2013AS,
Vanroose2006PRA,Horner2008PRL,Lee2010JPB}.
However, direct real-space {\color{black}solutions}
of \textcolor{black}{the} TDSE for systems with more than two electrons
remain a major challenge. 

To investigate  multi-electron dynamics in intense laser fields, 
the multiconfiguration time-dependent Hartree-Fock (MCTDHF) method has
been developed \cite{Zanghellini:2003,Zanghellini:2004,Kato:2004,Caillat:2005,Nest:2005a} in which the
time-dependent total 
%wave function is expanded in terms of the Slater determinants, 
wave function is given in the configuration interaction (CI) expansion,
\begin{eqnarray}\label{eq:mcscf}
 \Psi(t) = \sum_I \Phi_I(t) C_I(t),
\end{eqnarray}
where $\Phi_I(t)$ is a Slater determinant built from a given
number, $n$, of orbital functions $\{\psi_p(t)\}$ called occupied orbitals.
Both the CI coefficients $\{C_I\}$ and the orbitals are
simultaneously varied in time which allows 
%an accurate representation of the total wave function with 
{\color{black}to use} a considerably smaller number of orbitals than in a standard CI approach.
The conventional MCTDHF method is based on the full-CI expansion; 
the summation $I$ in Eq.~(\ref{eq:mcscf}) is taken over all
possible realizations of distributing $N$ electrons among the $2n$ spin orbitals
$\{\psi_p\}\otimes\{\uparrow,\downarrow\}$, where $\uparrow$
($\downarrow$) is the up- (down-)spin eigenfunction.
In this article we refer to the term time-dependent multi-configurational 
self-consistent field (TD-MCSCF) method in a broader context that involves the multiconfiguration wave function of the form
Eq.~(\ref{eq:mcscf}) without the restriction to the full-CI expansion.

{\color{black}Following} the first implementations of the MCTDHF method for one{\color{black}-}dimensional (1D) model
Hamiltonians \cite{Zanghellini:2003,Zanghellini:2004,Caillat:2005} and with a Gaussian basis set
\cite{Kato:2004}, three{\color{black}-}dimensional (3D) real-space implementations and
their applications have been reported by several authors
\cite{Jordan:2006,Kato:2008,Hochstuhl:2011,Haxton:2012,Haxton:2014}. 
%It has been also extended to the quantum mechanical treatment of nuclei \cite{}.
%
The 3D implementations have been most successfully applied to cases with
short wavelength (high photon energy) pulses. 
For example, Hochstuhl and Bonitz have
presented an application to atoms 
in spherical polar coordinates and simulated
%%%%expanded orbital functions for atoms with spherical
%%%%harmonics with the radial coordinate described with the
%%%%finite-element discrete variable representation (FEDVR) basis.
%%They applied the spherical-FEDVR implementation to 
two-photon ionization of helium \cite{Hochstuhl:2011}. 
Haxton and McCurdy have used spherical polar coordinates and prolate
spheroidal coordinates for atoms and diatomic molecules, respectively,
%with (FE)DVR basis used for radial dgrees of freedom, 
and simulated single-photon ionization of a Be atom and {\color{black}a} HF molecule \cite{Haxton:2012}.
They have also simulated X-ray core excitation and core ionization and
subsequent relaxation processes in NO molecules \cite{Haxton:2014}.
In contrast to perturbative processes, application of the MCTDHF method
to strong-field processes at longer wavelength (visible to infrared) and
{\color{black}high intensities} (peak intensities up to 10$^{15}$ 
W/cm$^2$) such as tunneling ionization and high-harmonic generation (HHG) are largely missing. 
A few exceptions include the work of Jordan {\it et al.} \cite{Jordan:2006} who analyzed molecular size effects on strong-field
ionization of model two electron systems, and the work of
Kato and Kono \cite{Kato:2008}, where single and double ionization of a hydrogen molecule
induced by a near-infrared (NIR) laser pulse were investigated.
The quantitative first-principles study of multi-electron dynamics 
in the long-wavelength high-intensity regime still remains a challenge.

%The difficulty of such applications is summarized in two points; 
%(i) the large CI dimension [expansion length of Eq.~(\ref{eq:mcscf})] required to accurately describe
%many-electron wave function, and (ii) large simulation volume and high
%spatial resolution required to represent electron motions with large
%quiver-amplitude and high kinetic energy.
%

One of the difficulties lies in the large CI
dimension defined as the expansion length in Eq.~(\ref{eq:mcscf}),
required to accurately describe many-electron wave functions. Within 
the full-CI based MCTDHF method the CI dimension, and therefore, the
computational cost increases factorially with the number of electrons $N$.
To overcome this limitation we have recently proposed the time-dependent
complete-active-space self-consistent-field (TD-CASSCF) method \cite{Sato:2013}.
Similar to the stationary CASSCF method of quantum chemistry it makes use of the decomposition into {\it core} and {\it active}
orbital subspaces. Accordingly, 
%The core orbitals are forced to be
%doubly occupied all the time, while the active orbitals are allowed
%general (0, 1, or 2) occupancies.
%Thus, the CASSCF wave function is written symbolically as
%\begin{subequations}
%\label{eq:casci_symbolic} 
%\begin{eqnarray}
%\Psi_{\rm CASSCF}(t) &:& \phi^2_1 \phi^2_2 \cdot\cdot\cdot
%\phi^2_{n_\textrm{c}} \label{eq:casci_symbolic_core} \\ &\times&
%\left\{
%\phi_{n_\textrm{c} + 1} \phi_{n_\textrm{c} + 2} \cdot\cdot\cdot \phi_{n_\textrm{c} + n_\textrm{a}}
%\right\}^{N_\textrm{a}}, \label{eq:casci_symbolic_active} 
%\end{eqnarray}
%\end{subequations}
%where factors (\ref{eq:casci_symbolic_core}) and
%(\ref{eq:casci_symbolic_active}) represent core- and active-subspaces,
%respectively, and the total wave function is properly antisymmetrized. 
%See Sec.~\ref{subsec:mcscf} for the rigorous definition.
%old The $n_\textrm{c}$ core orbitals describe $N_\textrm{c} = n_\textrm{c} / 2$
%old core electrons within the closed-shell wave function as in the time-dependent
%old Hartree-Fock (TDHF) method {\color{black}\cite{Kulander:1987}}, while the  $N_\textrm{a}=N-N_\textrm{c}$ active electrons are fully
%old correlated among $n_\textrm{a}=n-n_\textrm{c}$ active orbitals as in the
%old MCTDHF method. 
%
core electrons within the closed-shell wave
function are treated closely following the time-dependent Hartree-Fock (TDHF) method
{\color{black}\cite{Kulander:1987}}, while the active electrons are
fully correlated among active orbitals as in the MCTDHF method. 
%Note that the total wave function is properly antisymmetrized.
Whereas, in general, all the orbitals are varied in time, it is
possible to further split the core space {\color{black}into} time-independent frozen-core (FC) and time-dependent dynamical-core (DC) orbitals (see Fig.~\ref{fig:cas}).
With the decomposition into core and active orbitals, the TD-CASSCF method significantly reduces the CI
dimension without sacrificing the accuracy in the description of
multi-electron dynamics in long-wavelength high-intensity lasers.
{\color{black} 
The TD-CASSCF method is gauge invariant \cite{Ishikawa:2015} and size
extensive \cite{Helgaker:2002}. 
The fully correlated active space enables an accurate description of ionization processes 
including multichannel and multi-electron effects while dynamical-core
orbitals efficiently account for the field-induced core
polarization. 
}
More approximate and thus computationally even {\color{black}less demanding}
methods have also been developed \cite{Miyagi:2014b, Haxton:2015,
Sato:2015}, such as the time-dependent occupation-restricted multiple
active-space (TD-ORMAS) method \cite{Sato:2015}. 
See Ref.~\cite{Ishikawa:2015} for a broad
review of {\it ab initio} methods for multi-electron dynamics.

This paper reports on an efficient fully {\color{black}3D} implementation of the
TD-CASSCF method for atoms in the field of a linearly polarized laser pulse.
Simulations for long-wavelength high-intensity pulses involve a large simulation volume and high
spatial resolution to represent the electronic motion with large
quiver amplitudes and high kinetic energy, requiring a very large number $K$ of basis functions (or equivalently, grid points) 
for expanding {\color{black}the} orbital functions. We reduce the resulting computational cost and harness the
advantages of the TD-CASSCF method by introducing a gauge-invariant description of the frozen-core subspace allowing a velocity-gauge simulation, which is
known to be superior to the length gauge treatment for strong field phenomena {\color{black}\cite{Nurhuda:1998,Grum-Grzhimailo:2010}}. 
{\color{black}Our} implementation employs a spherical harmonics expansion of
orbitals with the radial coordinate discretized by {\color{black}a} finite-element discrete
variable representation (FEDVR) \cite{Rescigno:2000,McCurdy:2004,Schneider:2006,Schneider:2011}.
For the computationally most costly operation, the evaluation of the mean field, we use a Poisson solver thereby achieving linear scaling with $K$. A split-operator propagator is
developed with an efficient implicit method for {\it stiff} derivative
operators which drastically stabilizes the temporal propagation of orbitals. 
Combining these techniques makes it possible to take full advantage of
the TD-CASSCF method and permits benchmark calculations for atoms.
We present the HHG spectra for He, Be, and Ne atoms induced
by an intense NIR laser pulse {\color{black} and explore the effect of the electron correlation. Our results are converged with respect to the spatial
and temporal discretization.}
{\color{black}
This paper is organized as follows. 
In Sec.~\ref{sec:theory}, we present the equations of motion (EOMs)
for the TD-CASSCF method and introduce a gauge-invariant frozen-core treatment. 
Our implementation of the TD-CASSCF method for atomic systems is described in
Sec.~\ref{sec:implementation}, and
numerical applications are described in Sec.~\ref{sec:numerical}.
Section~\ref{sec:summary} concludes this work and discusses future prospects. 
In order to improve the readability of the manuscript, we have moved a considerable amount of technical details to Appendices \ref{app:tdmcscf} to \ref{app:fedvr}.
%%% required correction in case using frozen-core orbitals.
Hartree atomic units are used throughout unless otherwise noted.
}
\section{Method\label{sec:theory}}
\subsection{The system Hamiltonian\label{subsec:hamiltonian}}
We consider an atom with $N$ electrons exposed to 
a laser field linearly polarized in the $z$ direction. The Hamiltonian reads
\begin{eqnarray}\label{eq:ham_1q}
H = \sum_{i=1}^N h(\bm{r}_i,\bm{p}_i,t) +
\sum_{i=1}^N\sum_{j>i} U(\bm{r}_i, \bm{r}_j)
\end{eqnarray}
where $\bm{r}_i$ and $\bm{p}_i=-i\bm{\nabla}_i$ are the coordinate and canonical
momentum of the electron $i$,
%%% $i$'s electron,
with the one-body Hamiltonian
\begin{eqnarray}\label{eq:h0_1q}
\label{eq:h1e_1q}
h(t) &=& h_0 + V_{\textrm{ext}}(t),
\end{eqnarray}
and the electron-electron interaction
\begin{eqnarray}\label{eq:v2_1q}
U(\bm{r}_1,\bm{r}_2) &=& \frac{1}{r_{12}} = \frac{1}{|\bm{r}_1-\bm{r}_2|}.
\end{eqnarray}
The atomic Hamiltonian $h_0$ and 
the laser-electron interaction $V_{\rm ext}$ 
within the dipole approximation either in {\color{black}the} length gauge
(LG) or in {\color{black}the} velocity gauge (VG) are given by
\begin{eqnarray}\label{eq:vext_1q}
h_0(\bm{r},\bm{p}) = \frac{1}{2}p^2 + V_0(\bm{r}), \hspace{.5em}
V_0(\bm{r}) = -\frac{Z}{r}
\end{eqnarray}
\begin{subequations}\label{eqs:laser}
\begin{eqnarray}
V_{\rm ext}^{\textrm{LG}}(\bm{r},t) &=& \label{eq:laser_length}
E(t) z, \\
V_{\rm ext}^{\textrm{VG}}(\bm{p},t) &=& \label{eq:laser_length}
A(t) p_{z},
\end{eqnarray}
\end{subequations}
where 
%%%the polarization direction is chosen to be $z$-axis,
$V_0$ is the nuclear potential with $Z$ being the atomic number, and
$E(t)$ and $A(t)=-\int E(t)dt$ are the laser electric field and the vector
potential, respectively. 
We will use the second-quantized operators $\hat{h}$,
$\hat{U}$, and $\hat{H}$, etc{\color{black}.}, corresponding to those defined in
real space in Eqs.~(\ref{eq:ham_1q}) to (\ref{eqs:laser}). {\color{black}Explicit expressions for the operators in second quantization are given in
Appendix~\ref{app:tdmcscf}.}
%We will use the second-quantized operators $\hat{h}$,
%$\hat{U}$, and $\hat{H}$, etc, corresponding to those defined in the 
%real space in Eqs.~(\ref{eq:ham_1q}) to (\ref{eqs:laser}).
%The expression of operators in the second quantization is given in %Eqs.~(\ref{eq:ham}) to (\ref{eq:ham2e}) of 
%Appendix~\ref{app:tdmcscf}.
%%% The expression of the Hamiltonian %[Eqs.~(\ref{eq:ham_1q}) to (\ref{eqs:laser})]
%%% in the second quantization ($\hat{h}, $\hat{U}, \hat{H}, etc) is
%%% given in 
%%% %Eqs.~(\ref{eq:ham}) to (\ref{eq:ham2e}) of 
%%% Appendix~\ref{app:tdmcscf}.
%}

\subsection{TD-CASSCF method\label{subsec:tdcasscf}}
\begin{figure}[!b]
\centering
\includegraphics[width=1.0\linewidth,clip]{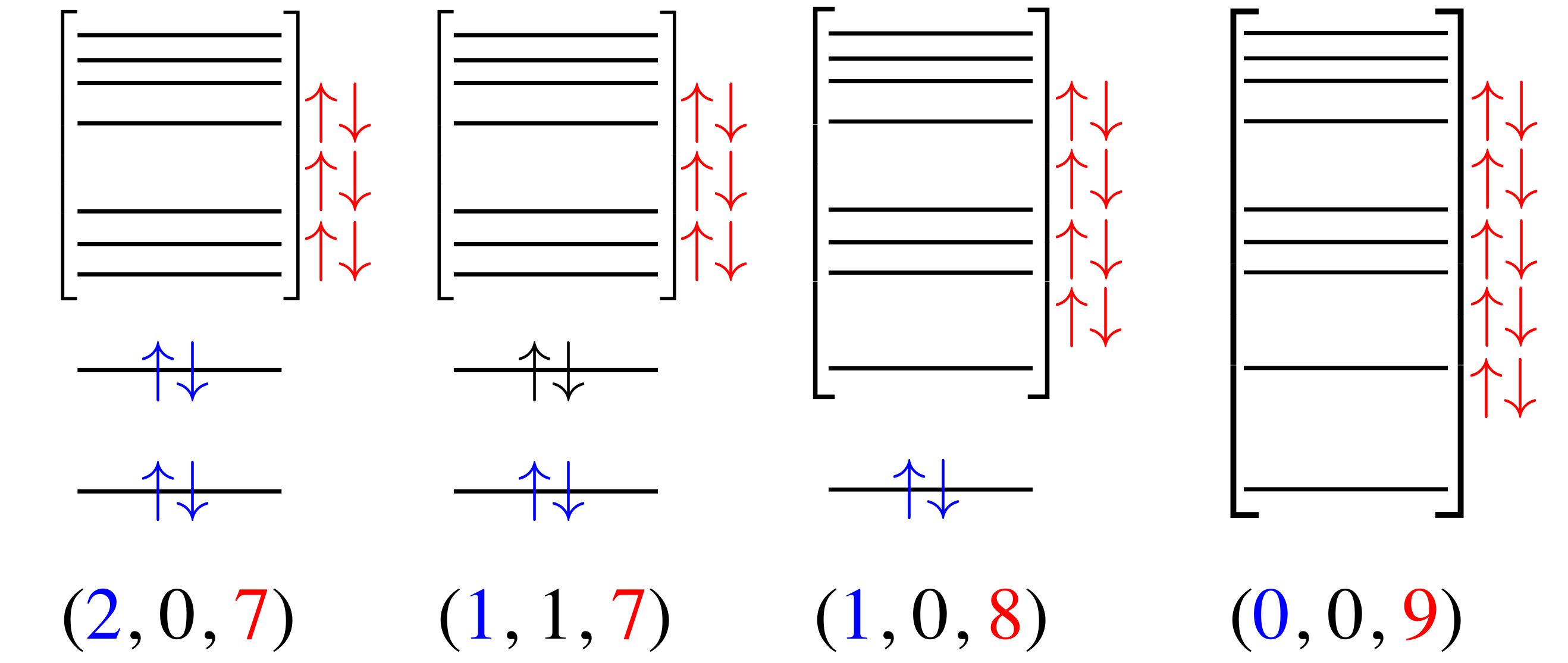}
\caption{\label{fig:cas} \color{black}Illustration of the TD-CASSCF
concept for a 10-electron system with 9 occupied orbitals. 
The up and down arrows represent electrons decomposed into
FC electrons (blue), DC electrons (black){\color{black},} and active electrons (red).
The horizontal lines represent occupied orbitals, classified into doubly occupied
FC and DC orbitals and active orbitals (bracketed). Three
examples, (2,0,7), (0,1,8), and (0,0,9), correspond to the
methods used for {\color{black}the} Ne atom in Sec.~\ref{subsec:numerical_ne}.
}%
\end{figure}
{\color{black}
In the TD-CASSCF method \cite{Sato:2013}, the $n$ occupied orbitals are separated into
$n_\textrm{c}$ doubly occupied core orbitals $\{\psi_i: i=1,2,\cdot\cdot\cdot,n_\textrm{c}\}$ and 
$n_\textrm{a}$ active orbitals $\{\psi_t: t=n_\textrm{c}+1,n_\textrm{c}+2,\cdot\cdot\cdot,n\}$,
with $n = n_\textrm{c} + n_\textrm{a}$. The core orbitals can be further
decomposed into $n_\textrm{fc}$ FC orbitals and $n_\textrm{dc}$ DC orbitals,
with $n_\textrm{c} = n_\textrm{fc} + n_\textrm{dc}$. Then the $N$-electron wave function is given by
%%% expanded with those Slater determinants
%%% including $n_\textrm{c}$ doubly occupied core orbitals,
\begin{eqnarray}\label{eq:casscf}
\Psi_\textrm{CAS} = \hat{A}\left[\Phi_\textrm{fc}\Phi_\textrm{dc}\sum_I \Phi_I C_I\right],
\end{eqnarray}
where $\hat{A}$ is the antisymmetrization operator, $\Phi_\textrm{fc}$ and $\Phi_\textrm{dc}$ are the closed-shell
determinants constructed with FC and DC orbitals, respectively, and
$\{\Phi_I\}$ are the determinants formed {\color{black}by} active orbitals.
In the following we will denote the level of the CAS approximation
employed in {\color{black}$\Psi_{\rm CAS}$}
by the integer triple  $(n_\textrm{fc}, n_\textrm{dc}, n_\textrm{a})$ with $n_\textrm{fc} + n_\textrm{dc} + n_\textrm{a}=n$.
The MCTDHF method corresponds to the special case $(0,0,n)$. As the formulation of the TD-CASSCF method involves 
different classes of orbitals{\color{black},} we introduce for clarity specific index sets: for core orbitals we will use $\{i,j\}$,
for active orbitals $\{t,u,v,w,x\}$ and for arbitrary orbitals (core and active) $\{o,p,q,r,s\}$. The FC and
DC orbitals are distinguished explicitly only when necessary.

{\color{black}
The equations of motion (EOMs) for the TD-CASSCF method has been derived based on
the time-dependent variational principle \cite{Frenkel:1934,
Lowdin:1972, Moccia:1973}, where the following action integral $S$,
\begin{eqnarray}\label{eq:action}
S=\int dt \langle\Psi|\hat{H}-i\frac{\partial}{\partial t}|\Psi\rangle,
\end{eqnarray}
is required to be stationary, i.e, $\delta S = 0$ for the variation of orbitals
$\{\psi_p\}$ and CI coefficients $\{C_I\}$. The form of the resulting EOMs {\color{black}is} not unique
but can be written in various equivalent {\color{black}ways} (see Appendix~\ref{app:oorot} for further details). Here we present 
the EOMs in the form particularly well-suited for the split-operator method applied below. The EOMs for the CI coefficients read
\begin{subequations}\label{eqs:eom_split}
\begin{eqnarray}
\label{eq:eom_split_cic}
i \dot{C}_I &=& \sum_J \langle \Phi_I|\hat{U}|\Phi_J\rangle C_J,
\end{eqnarray}
%{\color{black} where $\hat{U}$ is the {\color{black} electron-electron
%interaction operator [second term of  Eq.~(\ref{eq:ham_1q})]},}
and the EOMs of the orbitals are given by
\begin{eqnarray}
\label{eq:eom_split_orb}
i|\dot{\psi}_p\rangle &=& \hat{h}|\psi_p\rangle +\hat{Q} 
\hat{F} |\psi_p\rangle + \sum_q |\psi_q\rangle R^q_p,
\end{eqnarray}
\end{subequations}
where $\hat{Q}=1-\sum_p |\psi_p\rangle\langle\psi_p|$ is the projector
onto the orthogonal complement of the occupied orbital space. The
operator $\hat{F}$ is defined by  
\begin{align}\label{eq:fock2}
%{\color{black}
%\hat{F} |\psi_p\rangle = \sum_{oqsr} (D^{-1})_o^q P^{os}_{pr} \hat{W}^r_s |\psi_q\rangle,
%} \nonumber \\
{\color{black}
\hat{F} |\psi_p\rangle = \sum_{oqsr} (D^{-1})_p^o P^{qs}_{or} \hat{W}^r_s |\psi_q\rangle,
}
\end{align}
where {\color{black} $D$} and {\color{black} $P$} are the one- and two-electron reduced
density matrix (RDM){\color{black}, respectively
(see} Ref.~\cite{Sato:2013} for their definition and the simplification
due to the core-active separation), 
and 
\begin{eqnarray}\label{eq:meanfield}
W^p_q(\bm{r}_1) &=&
\int d{\bm{r}_2}
\frac{\psi^*_p({\bm{r}_2}) \psi_q({\bm{r}_2})}
{|\bm{r}_1 - \bm{r}_2|}.
\end{eqnarray}
is the matrix element for the Coulomb interaction. The matrix element $R^q_p$,
\begin{eqnarray}\label{eq:nonredundent}
%{\color{black}
%R^q_p \equiv i \langle\psi_p|\dot{\psi}_q\rangle-h^q_p 
%} \nonumber \\
{\color{black}
R^q_p \equiv i \langle\psi_q|\dot{\psi}_p\rangle-h^q_p
}
\end{eqnarray}
determines the components of the time derivative of orbitals
{\color{black}in the subspace spanned by the occupied orbitals}.
The elements within one subspace, i.e.{\color{black},} $R^i_j$ and $R^u_t$, can be
arbitrary Hermitian matrix elements and are set to zero $R^i_j=R^u_t=0$
in the present implementation \cite{Sato:2013}. Elements between the
core and active space are given by \cite{Sato:2013}, 
\begin{eqnarray}\label{eq:x_ca}
%{\color{black}
%R^t_i = \sum_u (2-D)^{-1}_{tu} \left(2F^u_i-\sum_v F^i_v D^v_u\right),
%}
\nonumber \\
{\color{black}
R^t_i = \sum_u (2-D)^{-1}_{tu} \left(2F^u_i-\sum_v D^u_v F^{i*}_v\right),
}
\end{eqnarray}
%\begin{eqnarray}\label{eq:x_ca}
%Y_{ti} &=& -i\sum_u \left(2-D\right)^{-1}_{tu} \left(2F^{\rm 2e}_{ui}-\sum_v F^{{\rm 2e}*}_{iv} D^v_u\right),
%\end{eqnarray}
%%% \begin{eqnarray}\label{eq:y_ca}
%%% Y_{ti} = \sum_u (2-D)^{-1}_{tu} \left(2F^{\rm 2e}_{ui}-\sum_v F^{{\rm 2e}*}_{iv} D^v_u\right),
%%% \end{eqnarray}
where 
%{\color{black}$F^i_v = \langle\psi_i | \hat{F} | \psi_v \rangle$}
{\color{black} $F^p_q = \langle\psi_p | \hat{F} | \psi_q \rangle$}.
The explicit solution of this set of EOMs (Eq.~\ref{eqs:eom_split}) in a spherical basis will be discussed below. 

One important aspect of the TD-CASSCF method to be addressed is the
preservation of gauge invariance. While the MCTDHF method is gauge
invariant from the outset, the inequivalent treatment of frozen-core and
active orbitals within the TD-CASSCF method requires the explicit
introduction of gauge phases. 

Previously \cite{Sato:2013,Sato:2015}{\color{black},} the FC orbitals {\color{black}were}
identified with the field-free orbitals. 
This choice is justified only for the length gauge (LG). In order to render the TD-CASSCF gauge-invariant and to allow for the use of the numerically convenient velocity gauge (VG) we introduce gauge-dependent frozen-core orbitals as
\begin{eqnarray}
\label{eq:fc_lg}
 |\psi_i(t)\rangle =
\left\{
\begin{array}{cc}
|\psi_i(0)\rangle            & {\rm (LG)}\\
e^{-iA(t)z}|\psi_i(0)\rangle & {\rm (VG)}\\
\end{array}
\right. (i \in \textrm{FC}).
%%% {\rm LG} &:&\hspace{1.em} |\psi_k(t)\rangle = |\psi_k(0)\rangle, \\
%%% {\rm VG} &:&\hspace{1.em} |\psi_k(t)\rangle = e^{-iA(t)z}|\psi_k(0)\rangle,
\label{eq:fc_vg}
\end{eqnarray}
Correspondingly the matrix element (Eq.~\ref{eq:nonredundent})
\begin{eqnarray} \label{eq:xkp}
%%% &
%%% X_{ip} = -X^*_{p i} =
%%% \left\{
%%% \begin{array}{cc}
%%% 0 & {\rm (LG)}\\
%%% iE(t)z_{i p} &  {\rm (VG)}\\
%%% \end{array}
%%% \right.,& \\
&
R^p_i = (R^i_p)^* = 
\left\{
\begin{array}{cc}
-h^p_i & {\rm (LG)}\\
-h^p_i - E(t)z^p_i &  {\rm (VG)}\\
\end{array}
\right.,&
%&(i \in \textrm{FC}, p \notin \textrm{FC}).&
\end{eqnarray}
where $i \in \textrm{FC}$ 
and $z^p_i=\langle\psi_p | z | \psi_i \rangle$.}

\section{Implementation for many-electron atoms\label{sec:implementation}}
\subsection{Spherical-FEDVR basis\label{subsec:spherical_fedvr}}
In our implementation of the TD-CASSCF method for many-electron atoms we adopt the so called spherical-FEDVR basis functions,
\begin{eqnarray}\label{eq:fedvr}
\chi_{klm}(r,\theta,\phi) = \frac{1}{r} f_k(r) Y_{lm}(\theta,\phi),
\end{eqnarray}
where $(r,\theta,\phi)$ are the spherical coordinates of $\bm{r}$,
$Y_{lm}$ is a spherical harmonic, and $f_k(r)$ are normalized radial-FEDVR basis functions \cite{Rescigno:2000,McCurdy:2004}. We divide
the radial coordinate $[0,R_{\rm max}]$ 
into $K_{\rm FE}$ elements %
with each element
supporting $K_{\rm DVR}$ local DVR functions. The radial
coordinate is thus discretized into 
%$K_\textrm{rad}=K_{\rm FE}K_{\rm DVR}$ 
{\color{black}
$K_\textrm{rad}=K_{\rm FE}K_{\rm DVR}-(K_{\rm FE}-1)$ 
}
grid points $\{r_k\}$ {\color{black}(the {\color{black}term} $K_{\rm
FE}-1$ takes into account the bridge functions
\cite{Rescigno:2000,McCurdy:2004} used for boundaries of neighboring finite elements)}, 
with integration weights $\{w^\textrm{rad}_k\}$ such that
\begin{eqnarray}\label{fedvr}
f_k(r_{k^\prime}) = \frac{\delta_{kk^\prime}}{\sqrt{w^\textrm{rad}_k}},
\end{eqnarray}
and $r_1=0$, $r_{K_\textrm{rad}}=R_{\rm max}$ (For further details, see Refs.~\cite{Hochstuhl:2011,Rescigno:2000,McCurdy:2004}). We use $m$-adapted orbitals with fixed magnetic quantum number $m_p$ (see Appendix \ref{app:m_const}). Thus, we expand $\psi_p$ as
\begin{eqnarray}\label{eq:orb_sphfedvr}
\psi_p(r,\theta,\phi,t) = \sum_{k=2}^{K_\textrm{rad}-1} \sum_{l=0}^{L_{\rm max}}
\chi_{klm_p}(r,\theta,\phi) \bm{\varphi}^{kl}_p(t),
\end{eqnarray}
where the first and last FEDVR basis function {\color{black}are} removed to ensure the boundary
condition that orbitals vanish at both edges of the simulation box
\cite{McCurdy:2004}. The spherical harmonics expansion is truncated
after the maximum angular momentum $L_{\rm max}$. In total, the number of basis functions in Eq.~(\ref{eq:orb_sphfedvr}) is $K = (K_\textrm{rad}-{\color{black} 2}) (L_{\rm max}+1)$.
The vector $\bm{\varphi}_p$ of expansion coefficients with elements
$\bm{\varphi}^{kl}_p$
is our working variable, with which the orbital EOM {\color{black}Eq.}~(\ref{eq:eom_split_orb}) reduces to a matrix equation,
\begin{eqnarray}\label{eq:eom_matrix}
\label{eq:eom_orb_matrix}
i\frac{d}{dt}{\bm{\varphi}}_{p} &=& \bm{h}\bm{\varphi}_{p}
+\bm{Q} \bm{F} \bm{\varphi}_{p} + \sum_q \bm{\varphi}_{q} R^q_p.
\end{eqnarray}
%where $\bm{F}_p$ represent the operator $\hat{F}_p$ defined in
%Eqs.~\ref{eq:fock}, and $\bm{P}=\sum_p\bm{\varphi}_p\bm{\varphi}^\dagger_p$.
%We use the bold type face to denote the spherical-FEDVR representation
%of operator $\hat{h}$ as $\bm{h}$ with elements 
%$\bm{h}^{klm}_{k^\prime l^\prime m^\prime} = \langle\chi_{klm}|\hat{h}|\chi_{k^\prime l^\prime m^\prime}\rangle$.
Here, the matrices %$\bm{Q}=\bm{1}-\sum_p\bm{\varphi}_p\bm{\varphi}^\dagger_p$,
$\bm{Q}$, $\bm{h}${\color{black},} and $\bm{F}$ represent operators $\hat{Q}$,
$\hat{h}${\color{black},} and $\hat{F}$ in the spherical-FEDVR basis {\color{black}[}Eq.~(\ref{eq:fedvr}){\color{black}]}. Explicit expressions for matrix elements of the one-electron operator, $\bm{h}=\bm{h}_0+\bm{V}_{\rm ext}$, {\color{black}are} given in Appendix~\ref{app:fedvr}.
%old The spherical-FEDVR representations of one electron operators are well documented in the literature (See, e.g,
%old Refs.~\cite{Kulander:1992, Grum-Grzhimailo:2010}). Nonzero matrix elements are
%%% Note that the suffix $kl$ in the left-hand side of the
%%% following equations is applied to the result of matrix-vector multiplication.
%
For the electron-electron interaction, we use the multipole expansion 
\begin{eqnarray}\label{eq:1/r12_multipole}
\frac{1}{r_{12}} &=& \sum_{l=0}^{L_{\rm ee}} \sum_{m=-l}^l
\frac{4\pi}{2l+1} \frac{r^l_<}{r^{l+1}_>} %\nonumber \\ &\times&
Y_{lm}(\theta_1,\phi_1)
Y^*_{lm}(\theta_2,\phi_2),
%%% Y_{lm}(\Omega_1)
%%% Y^*_{lm}(\Omega_2),
\end{eqnarray}
where $r_>$ ($r_<$) is the greater (smaller) of $r_1$ and ${r_2}$, 
$L_{\rm ee}$ is the highest multipole rank reached, $L_{\rm ee} \leq
2L_{\rm max}$, for the basis expansion of orbitals with $l \le L_{\rm max}$.
The two-electron part $\bm{F}\bm{\varphi}_p$ is evaluated by
(i) first calculating the mean field, Eq.~(\ref{eq:meanfield}), by
solving Poisson's equation,
\begin{eqnarray}\label{eq:poisson}
\nabla^2 W^p_q(\bm{r}) = -4\pi\psi^*_p(\bm{r})\psi_q(\bm{r}),
\end{eqnarray}
in the spherical-FEDVR basis of Eq.~(\ref{eq:fedvr}) \cite{McCurdy:2004,Hochstuhl:2011},
and (ii) performing the integrals weighted with RDMs through Eq.~(\ref{eq:fock2})
in the two-dimensional grid representation of coordinates $(r,\theta)$
with the known $\phi$ dependence analytically integrated out (see
Appendix~\ref{app:fedvr} for further details).
%%% \begin{eqnarray}\label{eqs:fock2_sph_3d}
%%% \hat{V}_t \psi_t(r,\theta,\phi) &=& \sum_{uvw}
%%% W^{v}_{w}(r,\theta,\phi) \psi_u (r,\theta,\phi) {\it \Gamma}^{uw}_{tv}.
%%% \end{eqnarray}
%%% It is, therefore, advantageous to device a reversible
%%% transformation between the angular momentum and real-space
%%% representations in Eq.~(\ref{eq:orb_sphfedvr}),
%%% with the known $\phi$ dependence $e^{im_p\phi}$ of the orbital $\psi_p$
%%% being integrated out. For this purpose, we use the Gauss-Legendre transformation,
%%% \begin{eqnarray}
%%% A(\theta_j) &=& \sum_l^L T_{jl} S_{l} 
%%% \equiv \sum_{l=0}^L P_{lm}(\cos\theta_j) S_{l},
%%% \end{eqnarray}
%%% \begin{eqnarray}
%%% S_{l} &=& \sum_{j} T^{-1}_{lj} A(\theta_j) %\int_{-1}^1 d(\cos\theta) P^*_{lm}(\cos\theta) f(\theta) \\
%%% \equiv \sum_{j=0}^L w^{\rm ang}_j P^*_{lm}(\cos\theta_j) A(\theta_j)
%%% \end{eqnarray}
%%% where $A(\theta)$ is an $L$-th order polynomial of $\cos\theta$, expanded
%%% by the associated Legendre polynomial $P_{lm}$. The $\cos(\theta_j)$
%%% and $w^{\rm ang}_j$ are the node and weight of the Gauss-Legendre quadrature.

\subsection{Split operator propagator}\label{subsec:split}
The orbital EOM Eq.~(\ref{eq:eom_matrix}) consists of both the linear (i.e., independent of
orbital functions and CI coefficients) term $\bm{h}$ and the nonlinear terms containing dependencies on the orbitals and 
the CI coefficients. Moreover, the linear part contains the stiff spatial-derivative operators while the 
{\color{black} nonlinear} part is non-stiff. We therefore introduce an effective split-operator algorithm that 
uses adapted propagators for  $\bm{h}$ and the {\color{black} nonlinear} part. We use a second-order
split method \cite{Lubich:2004,Caillat:2005,Koch:2010}, in which the propagation for the time interval $[t,t+\delta t]$
is performed as follows. First we solve the linear equation
\begin{eqnarray}\label{eq:split_first}
\frac{d}{dt}\bm{\varphi}_p &=& -i\bm{h}(t) \bm{\varphi}_p,
\end{eqnarray}
for $[t,t+\delta t/2]$ with initial values
$\{\bm{\bm{\varphi}}_p(t)\}$ to obtain $\{\bm{\bm{\varphi}}^\prime_p\}$, with CI
coefficients kept fixed, $C^\prime_I = C_I(t)$.
Then Eqs.~(\ref{eqs:eom_split}), for the {\color{black} nonlinear} part and the
CI coefficients
are solved for $[t,t+\delta t]$ with initial values $\{\bm{\varphi}_p^\prime, C^\prime_I\}$ to obtain
$\{\bm{\varphi}_p^{\prime\prime}, C^{\prime\prime}_I\}$. Finally 
Eq.~(\ref{eq:split_first}) is solved again for $[t+\delta t / 2,
t+\delta t]$ with initial values $\{\bm{\varphi}_p^{\prime\prime}\}$ and CI coefficients kept fixed to obtain
$\{\bm{\varphi}_p(t+\delta t), C_I(t+\delta t)=C^{\prime\prime}_I\}$.
For the stiff Eq.~(\ref{eq:split_first}) we adopt the Crank-Nicolson method, 
\begin{eqnarray}\label{eq:crank_nicolson}
\bm{\varphi}_p(t+\delta t/2) &=&
\frac{1-i\bm{h}(t+\delta t/4)\delta t/4}
     {1+i\bm{h}(t+\delta t/4)\delta t/4}
\bm{\varphi}_p(t).
\end{eqnarray}
%%% with $\bm{h}^{\rm 1e}$ evaluated at the midpoint of the time interval to have second order accuracy. 
The right hand side is evaluated by the matrix iteration method \cite{Nurhuda:1998},
in which the inverse operator is expanded as
\begin{eqnarray}
\left(1+i\bm{h}\delta t^\prime\right)^{-1} 
&=&
\left\{\left(1+i\bm{h}_0\delta t^\prime\right)
\left(1 + \frac{i\bm{V}_\textrm{ext}\delta t^\prime}{1+i\bm{h}_0\delta t^\prime} \right)
\right\}^{-1} \nonumber \\ 
&=&
\left(1 + \frac{i\bm{V}_\textrm{ext}\delta t^\prime}{1+i\bm{h}_0\delta t^\prime}\right)^{-1}
\frac{1}{1+i\bm{h}_0\delta t^\prime} \nonumber \\
&=&
\sum_{m=0}^\infty \left(
-\frac{i\bm{V}_\textrm{ext}\delta t^\prime}{1+i\bm{h}_0\delta t^\prime}
\right)^m \!\!\!\! \frac{1}{1+i\bm{h}_0\delta t^\prime}, 
\end{eqnarray}
with $\delta t^\prime = \delta t/4$.
%%% \begin{eqnarray}
%%% \left(1+i\bm{h}^{\rm 1e}\delta t/4\right)^{-1} 
%%% &=&
%%% \left\{\left(1+i\bm{h}_0\delta t/4\right)
%%% \left(1 + \frac{i\bm{V}^\textrm{ext}\delta t/4}{1+i\bm{h}_0\delta t/4} \right)
%%% \right\}^{-1} \nonumber \\ 
%%% &=&
%%% \left(1 + \frac{i\bm{V}^\textrm{ext}\delta t/4}{1+i\bm{h}_0\delta t/4}\right)^{-1}
%%% \frac{1}{1+i\bm{h}_0\delta t/4} \nonumber \\
%%% &=&
%%% \sum_{m=0}^\infty \left(
%%% -\frac{i\bm{V}^\textrm{ext}\delta t/4}{1+i\bm{h}_0\delta t/4}
%%% \right)^m \!\!\!\! \frac{1}{1+i\bm{h}_0\delta t/4}, 
%%% \end{eqnarray}
This defines an iterative procedure,
\begin{eqnarray}
\bm{\varphi}_p(t+\delta t/2) \approx \sum_{j=0}^{N_\textrm{itr}} \bm{f}_j,
\end{eqnarray}
\begin{eqnarray}
\left(1+i\bm{h}_0\delta t^\prime\right)\bm{f}_0 &=& \left(1-i\bm{h}\delta t^\prime\right) \bm{\varphi}_p(t), \\
\left(1+i\bm{h}_0\delta t^\prime\right)\bm{f}_j &=& -i\bm{V}_\textrm{ext}\delta t^\prime \bm{f}_{j-1}
\hspace{1.0em} (j \geq 1).
\end{eqnarray}
This method has been found to be quite efficient for
ionization dynamics of atomic hydrogen under the presence of an intense,
long-wavelength laser field \cite{Grum-Grzhimailo:2010}.
For the nonlinear, but nonstiff part on the right-hand side of Eq.~(\ref{eq:eom_split_orb}), 
we use the fourth-order Runge-Kutta propagator. 

\begin{figure}[!t]
\centering
\includegraphics[width=0.95\linewidth,clip]{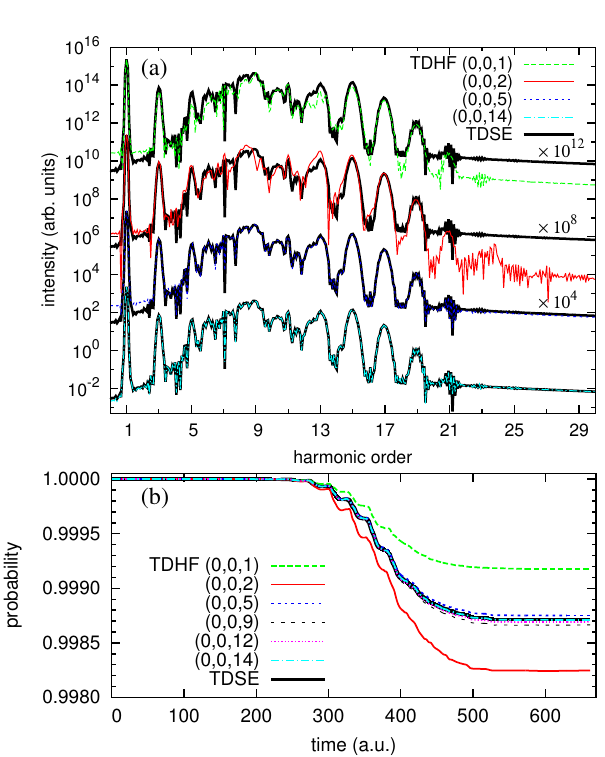}
\caption{\label{fig:he_w400}(a) High harmonic generation spectrum of a He
atom exposed to a visible laser pulse with a wavelength {\color{black}of} 400 nm and an intensity
of 4$\times$10$^{14}$ W/cm$^2$. (b) The probability to find both electrons within a sphere of radius $R_0=20$ a.u. Results of the TD-CASSCF (or MCTDHF) method with different number of active orbitals $(0,0,n)$ are compared to the result of the exact TDSE method. $(0,0,1)$ corresponds to TDHF.
}%
\end{figure}
\section{Numerical results for HHG in many-electron atoms\label{sec:numerical}}
In this section, we present numerical applications of the implementation of the TD-CASSCF method 
to many-electron atoms described in the previous section.
In all simulations reported below, the initial state is taken as the ground state
obtained through imaginary time propagation of the EOMs. We assume a laser field of the
following form:
\begin{eqnarray}\label{eq:laser}
E(t) = E_0 \sin(\omega_0 t) \sin^2\left(\pi \frac{t}{\tau}\right),
\hspace{.5em} 0 \leq t \leq \tau, 
\end{eqnarray}
with total duration $\tau$,
peak intensity $I_0=E_0^2$, wavelength
$\lambda=2\pi/\omega_0$, and period $T=2\pi/\omega_0$.
{\color{black}
For all simulations presented a uniform {\color{black}finite} element
length of $d_{\rm FE}$ is used. The code allows, however, variable
element sizes for the radial-FEDVR basis, if needed. 
}

\subsection{Helium\label{subsec:numerical_he}}
%\subsubsection{Comparison with TDSE results\label{subsubsec:numerical_he_w400}}
\begin{figure}[!b]
\centering
\includegraphics[width=0.95\linewidth,clip]{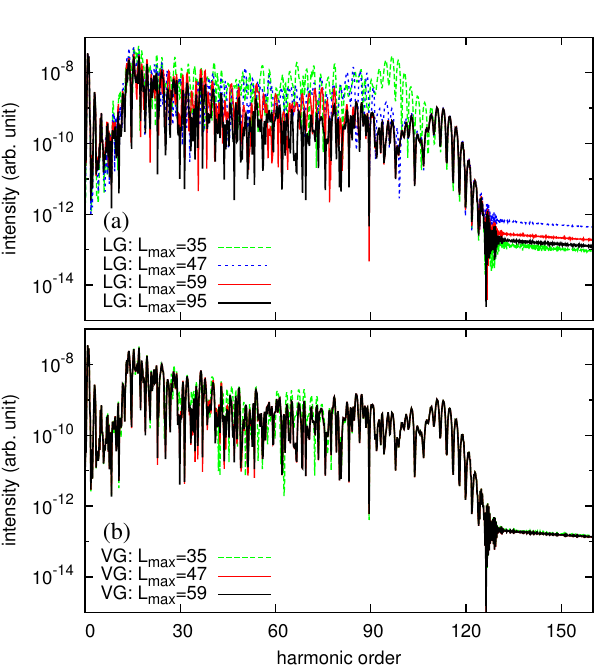}
\caption{\label{fig:he_hhg_w800_L1}HHG spectra of He exposed to an IR laser pulse
with a wavelength of 800 nm and an intensity of 8$\times$10$^{14}$
W/cm$^2$. Results for {\color{black} (a)} length gauge (LG) and {\color{black}(b)} velocity gauge (VG) simulations
with the TD-CASSCF (or MCTDHF) method with fixed orbital indices $(0,0,5)$ as a 
function of maximum angular momentum $L_{\rm max}$ in the orbital expansion.
}%
\end{figure}
First we simulate a helium atom subject to a field of 400~nm wavelength, 4.0$\times$10$^{14}$ W/cm$^2$ intensity, and {\color{black} total duration of 12 periods $T$ of the optical field ($\tau = 12T$)}. For this system, a direct exact numerical solution of the time-dependent {\color{black} Schr\"odinger} equation (TDSE) 
is possible and serves as a benchmark for the present TD-CASSCF code.
Note that for this two{\color{black}-}electron system, the TD-CASSCF method reduces to TDHF for the orbital 
choice ($n_\textrm{fc}=0,n_\textrm{dc}=0,n_\textrm{a}=1$) 
or to the MCTDHF method for ($n_\textrm{fc}=0,n_\textrm{dc}=0,n_\textrm{a}>1$).
The TDSE simulations are performed using the code developed
at Vienna University of Technology
\cite{Feist2009PRL,Pazourek2011PRA,Schneider:2011}, in which 
the six-dimensional two electron wave function is expanded with
coupled spherical harmonics, and the radial coordinates are discretized
with a FEDVR basis. The code is highly optimized and massively
parallelized enabling the large scale simulations as presented below.
Here we aim at a rigorous comparison between the TD-CASSCF (or MCTDHF) method
and the TDSE results under otherwise identical simulation conditions. 

We use the radial-FEDVR basis with {\color{black} $K_{\rm FE}=80$}, $d_{\rm FE}=4.0$,
$K_{\rm DVR}=11$, and $R_{\rm max}=320$, and spherical harmonics
expansions with $L_{\rm max}=L_{\rm ee}=13$.
%%% We used 80 finite elements of length 4.0 a.u 
%%% up to the radial box
%%% boundary at 320 a.u, with each element including 11 local DVR
%%% functions. 
To avoid a reflection at the simulation box boundary,
orbital functions on the radial grid points $r_k > 256$ are masked
after each time propagation by a $\cos^{1/4}$ function.
%%% The maximum angular momenta for expanding
%%% orbitals ($L_{\rm max}$) and Coulomb operators ($L_{\rm ee}$) are set $L_{\rm max}=L_{\rm ee}=13$.
Both TDSE and MCTDHF simulations employ the velocity gauge.
Several time-step sizes are tested (from $20000$ to $40000$ steps per
cycle) for the MCTDHF method. The TDSE simulation uses the short iterative
Lanczos propagator % with the maximum Krylov space dimension of 10 and
with self-adaptive time-step sizes.
We have carefully checked that the spatial and temporal
resolutions used are sufficient for reaching fully converged
results both for TDSE and MCTDHF simulations.
\begin{figure}[!b]
\centering
\includegraphics[width=0.95\linewidth,clip]{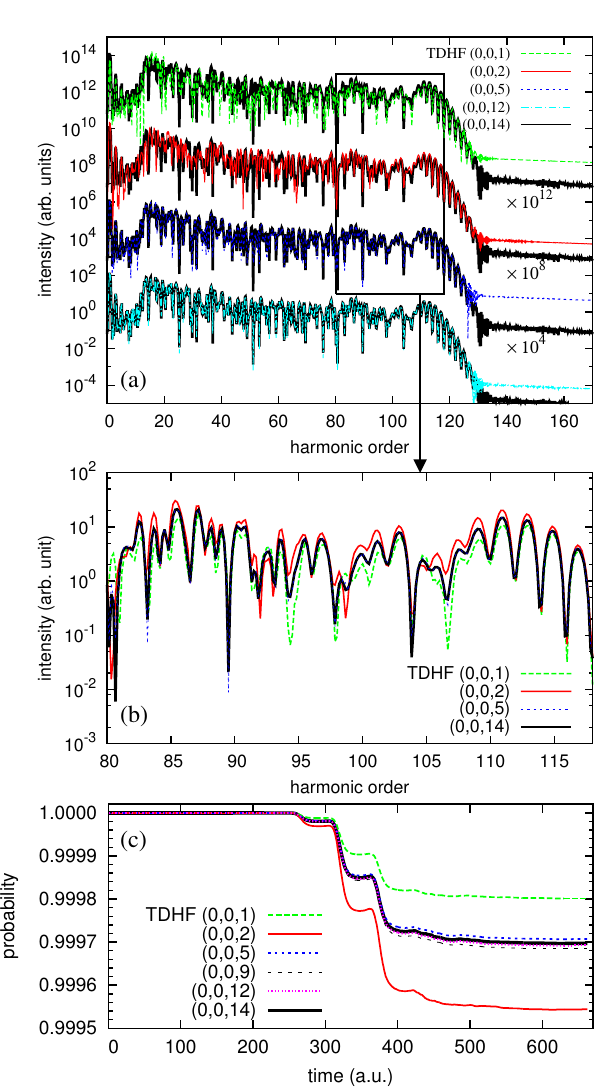}
\caption{\label{fig:he_w800_hhg_p0}(a) HHG spectra of He exposed to an IR laser pulse
with a wavelength of 800 nm and an
intensity of 8$\times$10$^{14}$ W/cm$^2$, (b) close-up of the region
 between the 80$^{\rm th}$ and the 118$^{\rm th}$ harmonic, and (c) the
 probability to find both electrons within a sphere of radius $R_0 = 20$
 a.u. Results of the TD-CASSCF (or MCTDHF) method are obtained with
 different number $n$ of active orbitals $(0,0,n)$ and maximum angular
 momentum 
{\color{black}
$L_{\rm max}=59$.
}
}%
\end{figure}
\begin{figure}[!t]
\centering
\includegraphics[width=0.95\linewidth,clip]{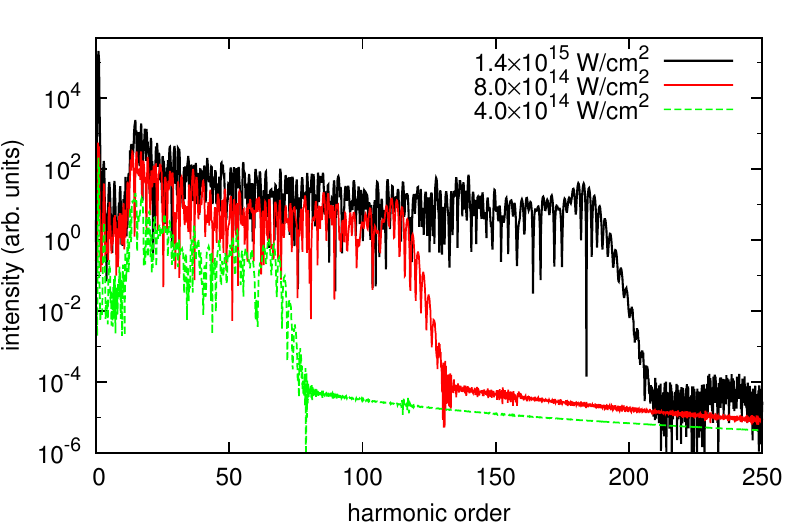}
\caption{\label{fig:he_w800_hhg_fint}HHG spectra of He exposed to an IR laser pulse
with a wavelength of 800 nm and varying intensities of 4, 8, and
14$\times$10$^{14}$ W/cm$^2$, obtained with the TD-CASSCF (or MCTDHF)
method with orbital indices $(0,0,14)$ and maximum angular momentum of $L_{\rm max}=71$.}
\end{figure}

The HHG spectra [Fig.~\ref{fig:he_w400}~(a)] calculated as the modulus squared of the Fourier transform of the
dipole acceleration 
[using the Ehrenfest expression Eq.~(\ref{eq:ehrenfest_acc}) below]
%%% We compare results of MCTDHF with one, two, five,
%%% and fourteen orbitals with that of TDSE simulation. 
display increasing agreement between MCTDHF and TDSE results 
{\color{black}as} the number of orbitals {\color{black} is increased}.
%%% get better in increasing the number of orbitals. 
The MCTDHF with $n \geq 5$
essentially reproduces the TDSE spectrum, and in particular, with $n=14$
the agreement is almost perfect over a dynamical range of 5 orders of magnitude. Likewise, the probability of finding 
both two electrons inside a sphere of radius $R_0 = 20$ a.u. (referred to as survival probability hereafter) measuring the temporal evolution of the complement to ionization or formation of Rydberg states [Fig.~\ref{fig:he_w400}~(b)] yields close agreement for $(0,0,n)$ with $n>2$. For $n=14$ the MCTDHF results become indistinguishable from TDSE. Note, however, significant differences to the TDHF limit pointing to the crucial role of correlations included in the MCTDHF method but neglected by TDHF.

Next we proceed to the convergence test as a function of $L_{\rm max}$
$(=L_{\rm ee})$ for both LG and VG. We consider now the
longer-wavelength and higher-intensity regime {\color{black}($\lambda=$ 800 nm,
$I_0=$ 8$\times$10$^{14}$ W/cm$^2$, and $\tau=6T$)}, for which convergence of the exact
TDSE code is still difficult to achieve. We reduce the FEDVR-element
size to $d_{\rm FE}$ = 2.0 in order to improve the representation of
higher-energy electrons and fix the orbital set to $(0,0,5)$. The HHG
spectrum converges for VG (Fig.~\ref{fig:he_hhg_w800_L1}) much faster
than for LG. While in the VG $L_{\rm max}=47$ suffices to reach overall good agreement with the fully converged results, angular momenta up to $L_{\rm max}\approx 90$ are needed in the LG for comparable accuracy. The convergence of the survival probability and of the harmonic spectrum as a function of $n$ (Fig.~\ref{fig:he_w800_hhg_p0}) is fairly rapid. Slight underestimation ($n=1$) and
overestimation ($n=2$) of harmonic intensities {\color{black}as} seen
[Fig.~\ref{fig:he_w800_hhg_p0}~(b)] 
in the upper plateau region,
%%% near the cutoff of the plateau,
are closely related to the underestimation ($n=1$) and overestimation
($n=2$) of the ionization [Fig.~\ref{fig:he_w800_hhg_p0}~(c)].
%%% This point will be revised below in Sec.~\ref{subsec:numerical_be}.
%
Finally, Fig.~\ref{fig:he_w800_hhg_fint} presents HHG
spectra for the wavelength 800 nm and three different intensities, 4,
8, and 14 $\times$ 10$^{14}$ W/cm$^2$. The results are converged, up to
the cutoff frequencies, in radial resolution ($d_{\rm FE}=2$),
angular resolution ($L_{\rm max}=L_{\rm ee}=71$), temporal resolution
(20000 steps per cycle), 
% {\color{black}
% electron correlation is fully accounted for at $n=14$. 
% }
{\color{black}
and number of orbitals ($n=14$). 
}
We stress the importance of {\color{black}the} VG implementation{\color{black},} with LG 
the convergence with respect to $L_{\rm max}$ was hardly reachable
%%% the convergence was hardly possible with respect to $L_{\rm max}$ 
for the highest intensity. The present simulation for helium demonstrates the accuracy of the present TD-CASSCF implementation by both the excellent agreement with the TDSE (where available) and the convergence with respect to all parameters characterizing the CASSCF orbital expansion. 

\subsection{Beryllium\label{subsec:numerical_be}}
%\subsubsection{Proper calculation of properties using frozen-core orbitals}

Beryllium is the first test case for the decomposition into frozen- and dynamical-core orbitals as well as active orbitals available in the TD-CASSCF method. Extraction of the harmonic spectrum via the expectation value of the dipole acceleration using {\color{black}the} Ehrenfest theorem requires additional care in the presence of frozen orbitals.  
Using the Ehrenfest theorem and {\color{black}the} canonical
commutation relation $[\hat{z},\hat{p}_z]=i$, the dipole moment,
dipole velocity, and dipole acceleration are given by 
\begin{eqnarray}\label{eq:ehrenfest_z}
\langle \hat{z} \rangle = \langle\Psi|\hat{z}|\Psi\rangle,
\end{eqnarray}
\begin{subequations}\label{eq:ehrenfest_velacc} 
\begin{eqnarray}\label{eq:ehrenfest_vel}
\frac{d}{dt}\langle \hat{z} \rangle = \langle\Psi|\hat{p}_z|\Psi\rangle,
\end{eqnarray}
and
\begin{eqnarray}\label{eq:ehrenfest_acc}
\frac{d^2}{dt^2}\langle \hat{z} \rangle = -
\langle\Psi|\left(
\frac{\partial \hat{V}_0}{\partial z}+
\frac{\partial \hat{V}_{\rm ext}}{\partial z}
\right)|\Psi\rangle,
\end{eqnarray}
\end{subequations}
respectively.
%%% Naively applying the same expression to the TD-MCSCF wave function with
%%% FC orbitals just replaces the expectation values for the total
%%% wave function in Eqs.~(\ref{eq:ehrenfest_z})-(\ref{eq:ehrenfest_acc})
%%% with those for the active wave function, 
%%% reflecting the very fact that the core wave function is frozen.
%%% \begin{eqnarray}\label{eq:ehrenfest_z_fc}
%%% \langle \hat{z} \rangle = \langle\Psi_\textrm{a}|\hat{z}|\Psi_\textrm{a}\rangle,
%%% \end{eqnarray}
%%% \begin{eqnarray}\label{eq:ehrenfest_vel_fc}
%%% \frac{d}{dt}\langle \hat{z} \rangle = \langle\Psi_\textrm{a}|\hat{p}_z|\Psi_\textrm{a}\rangle,
%%% \end{eqnarray}
%%% \begin{eqnarray}\label{eq:ehrenfest_acc_fc}
%%% \frac{d^2}{dt^2}\langle \hat{z} \rangle = -
%%% \langle\Psi_\textrm{a}|\left(
%%% \frac{\partial \hat{V}_0}{\partial z}+
%%% \frac{\partial \hat{V}_{\rm ext}}{\partial z}
%%% \right)|\Psi_\textrm{a}\rangle,
%%% \end{eqnarray}
%%% where Eqs.~(\ref{eq:ehrenfest_z})-(\ref{eq:ehrenfest_acc}) are simply
%%% replaced with expectation values of the active wave function only, 
%%% in Eqs.~(\ref{eq:ehrenfest_z_fc})-(\ref{eq:ehrenfest_acc_fc}),
%%% reflecting the very fact that the core wave function is frozen.
Equations~(\ref{eq:ehrenfest_z}) {\color{black}and} (\ref{eq:ehrenfest_velacc}) hold for the solutions $|\Psi\rangle$ of the TDSE as well as the TD-MCSCF. However, in the presence of frozen orbitals, Eqs.~(\ref{eq:ehrenfest_velacc}) must be modified to (for details see  Appendix~\ref{app:ehrenfest}),
%%% this is {\it not} a correct procedure to
%%% evaluate time derivatives of an expectation value.
%%% The correct expression is derived in Appendix~\ref{app:ehrenfest}, which adds 
%%% a correction term $\Delta$ given in Eq.~(\ref{eq:gbf_fc}),
\begin{subequations}\label{eq:ehrenfest_fc_velacc} 
\begin{eqnarray}\label{eq:ehrenfest_fc_vel}
\frac{d}{dt}\langle \hat{z} \rangle_\textrm{fc} = \langle\Psi|\hat{p}_z|\Psi\rangle + \Delta(\dot{z}),
%\frac{d}{dt}\langle \hat{z} \rangle = \langle\Psi|\hat{p}_z|\Psi\rangle + \Delta(\dot{z}),
\end{eqnarray}
\begin{eqnarray}\label{eq:ehrenfest_fc_acc}
\frac{d^2}{dt^2}\langle \hat{z} \rangle_\textrm{fc} = -
\langle\Psi|\left(
\frac{\partial \hat{V}_0}{\partial z}+
\frac{\partial \hat{V}_{\rm ext}}{\partial z}
\right)|\Psi\rangle +
\Delta(\dot{p}_z),
\end{eqnarray}
\end{subequations}
where the additional term $\Delta$ is defined by Eq.~(\ref{eq:gbf_fc}).
\begin{figure}[!b]
\centering
\includegraphics[width=0.9\linewidth,clip]{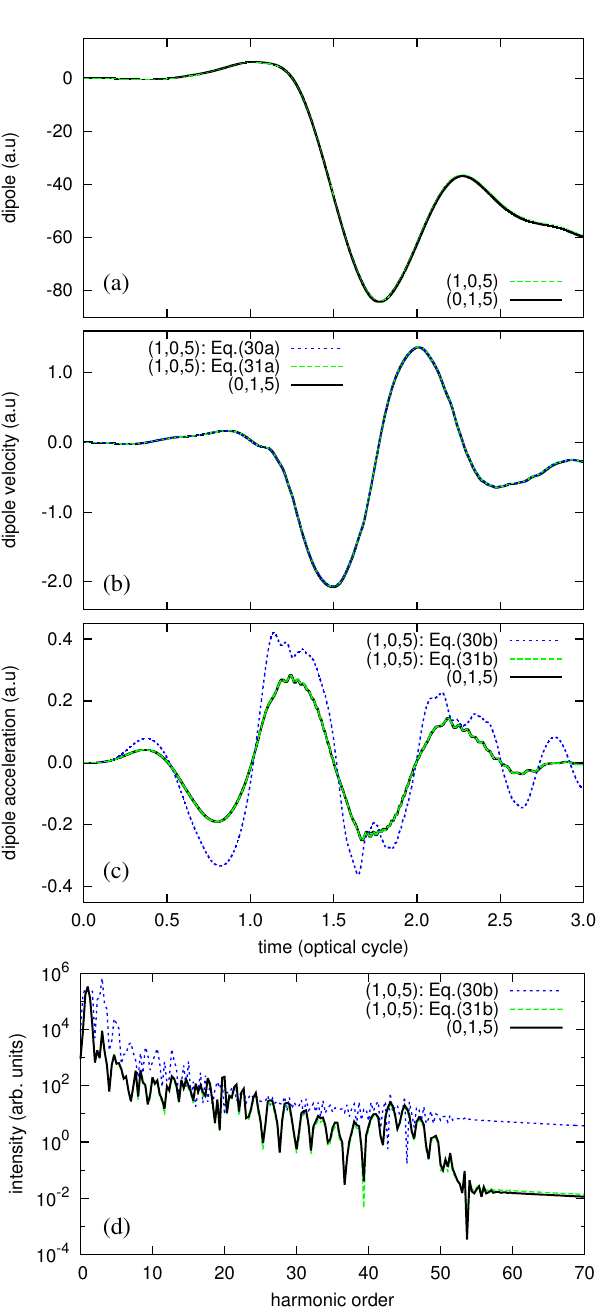}
\caption{\label{fig:be_tdhf}
%Time evolution of dipole moment (a),dipole velocity (b), and dipole acceleration (c), and HHG spectra (d) of Be exposed to an IR
{\color{black}
Time evolution of (a) the dipole moment, (b) the dipole velocity, {\color{black}and} (c) the dipole acceleration and (d) the HHG spectra 
} of Be exposed to an IR
laser pulse with a wavelength of 800 nm and an intensity of
3.0$\times$10$^{14}$ W/cm$^2$, obtained with the TD-CASSCF 
method with one dynamical-core orbital $(n_\textrm{fc}, n_\textrm{dc}, n_\textrm{a})=(0,1,5)$ {\color{black}or} one frozen-core orbital $(1,0,5)$ and a maximum angular momentum of $L_{\rm max}=47$.
}%
\end{figure}
The appearance of {\color{black}this} correction term in the presence of frozen orbitals can be qualitatively easily understood as follows: Focusing at the moment on the equations for the force [Eqs.~(\ref{eq:ehrenfest_acc}),~(\ref{eq:ehrenfest_fc_acc})], the dipole acceleration is given by the expectation value $\langle -\frac{\partial H}{\partial z} \rangle$ of the total force on the electronic system. It consists of the laser acting on the active {\color{black} electrons}, $f_{\rm la}$, the laser acting on the core electrons, $f_{\rm lc}$, the nuclear force acting on active, $f_{\rm na}$, and core electrons, $f_{\rm nc}$, and the inter-electronic forces, $f_{\rm ac}$ and $f_{\rm ca}$. With the action-reaction law, $f_{\rm ac}=-f_{\rm ca}$, the total force becomes
\begin{equation}\label{eq:force_ehrenfest}
f = (f_{\rm na} + f_{\rm nc}) + (f_{\rm la} + f_{\rm lc}),	
\end{equation}
which corresponds to the Ehrenfest expression Eq.~(\ref{eq:ehrenfest_acc}).
However, for the case of frozen-core orbitals, we have to include an additional ``binding force'' $f_{\rm b}$ to render the frozen-core immobile,
\begin{eqnarray} \label{eq:force_ehrenfest2}
f = (f_{\rm na} + f_{\rm nc}) + (f_{\rm la} + f_{\rm lc})+ f_{\rm b}.	
\end{eqnarray}
The ``binding force'' $f_{\rm b}$ must cancel all forces acting on frozen-core electrons,
\begin{eqnarray} \label{eq:force_ehrenfest3}
f_{\rm b} = - f_{\rm nc} - f_{\rm lc} - f_{\rm ac}.
\end{eqnarray}
Consequently, the effective force in the presence of a frozen core becomes 
\begin{eqnarray} \label{eq:force_fc}
f_{\rm eff} %&=& f_{\rm na} + f_{\rm la} - f_{\rm ac} \\
&=& (f_{\rm na} + f_{\rm ca}) + f_{\rm la},
\end{eqnarray}
i.e.{\color{black},} the inter-electronic force no longer cancels. The
correction term $\Delta(\dot{p}_z)$ represents the binding force $f_{\rm
b}$ [see Eq.~(\ref{eq:gbf_fc_approx_p})], and
Eq.~(\ref{eq:ehrenfest_fc_acc}) is the proper quantum
{\color{black}expression} for the effective force
[Eq.~(\ref{eq:force_fc})]. 
Use of the correct expression Eq.~(\ref{eq:ehrenfest_fc_acc}) is essential;
otherwise the missing binding force 
would lead to an incorrect evaluation of the dipole acceleration,
%%% prevents physically correct calculation of the dipole acceleration, 
and therefore of the HHG spectrum. This is illustrated in Fig.~\ref{fig:be_tdhf} showing the time
evolutions of (a) the dipole moment, (b) the dipole velocity, and (c) the dipole
acceleration of Be irradiated by 
% a laser pulse with a wavelength of 800 nm and an intensity of
% 3$\times$10$^{14}$ W/cm$^2$, 
{\color{black}
an intense NIR laser pulse ($\lambda=800$ nm, $I_0=4.0\times 10^{14}$ W/cm$^2$, and $\tau = 3T$)
}
calculated with the 
TD-CASSCF method either with {\color{black}the} DC [$(n_\textrm{fc}, n_\textrm{dc},
n_\textrm{a})=(0,1,5)$] or {\color{black}the} FC [$(n_\textrm{fc}, n_\textrm{dc},
n_\textrm{a})=(1,0,5)$] treatment of the initial $1s$ orbital.
For the FC case, we compare the standard Ehrenfest expression, Eq.~(\ref{eq:ehrenfest_velacc}),
and the one with the correction term, Eq.~(\ref{eq:ehrenfest_fc_velacc}).
The dipole moment, velocity, and acceleration
obtained with DC and FC simulations [Eqs.~(\ref{eq:ehrenfest_fc_velacc})]
agree very well {\color{black} with} each other, suggesting that FC is a good
approximation for beryllium. 
The FC velocities calculated with Eq.~(\ref{eq:ehrenfest_velacc}) and
Eq.~(\ref{eq:ehrenfest_fc_velacc}) are similar to each other, reflecting
the fact that $\Delta(\dot{z}) \approx 0$ as demonstrated in
Appendix~\ref{app:ehrenfest}. However, the FC acceleration evaluated by
the standard Ehrenfest expression, Eq.~(\ref{eq:ehrenfest_acc}),
fails. Only when the correction term $\Delta(\dot{p}_z)$ is included
[Eq.~(\ref{eq:ehrenfest_fc_acc})] excellent agreement between the FC and
DC treatment is achieved. 
The point to be emphasized is that it is not
the FC approximation itself
{\color{black}
but the evaluation of the {\color{black} nonlinear} response via the standard Ehrenfest expression, Eq.~(\ref{eq:ehrenfest_acc}), that fails.}
%{\color{black}
%but the use of the standard Ehrenfest expression, Eq.~(\ref{eq:ehrenfest_acc}), 
%that leads to the wrong {\color{black} nonlinear} response.
%}
{\color{black}
This observation has implications also for other approximation schemes in which
the core is kept frozen, most notably for TDSE solutions of
many-electron systems in the single-active electron (SAE) approximation.
%This observation also implies
%corrections for other approximation schemes in which the core is kept
%frozen, most notably for TDSE solutions of many-electron systems in the
%single-active electron (SAE) approximation. 
For example, Gordon {\it et
al.} \cite{Gordon2006PRL}
have argued that one should use Eq.~(\ref{eq:force_ehrenfest}) [Eq.~(10) of
Ref.~\cite{Gordon2006PRL}] rather than
Eq.~(\ref{eq:force_fc}) [equivalent to (the second time
derivative of) Eq.~(6) of Ref.~\cite{Gordon2006PRL}]. 
The present results suggest the opposite, i.e., that
acceleration in screened effective potential [Eq.~(\ref{eq:force_fc})]
rather than bare nuclear potential [Eq.~(\ref{eq:force_ehrenfest})] should be used.
}
%{\color{red}
%This observation has implications also for other approximation schemes in which
%the core is kept frozen, most notably for TDSE solutions of
%many-electron systems in the single-active electron (SAE) approximation. For 
%example, Gordon {\it et al.} \cite{Gordon2006PRL} have argued that one should 
%use Eq. (32) [Eq. (10) of Ref.~\cite{Gordon2006PRL}] rather than
%Eq. (35) [equivalent to (the second time derivative of) Eq. (6) of
%Ref.~\cite{Gordon2006PRL}]. The present results 
%suggest that acceleration in screened rather than bare potential should
%be used.
%}

\begin{figure}[!t]
\includegraphics[width=0.95\linewidth,clip]{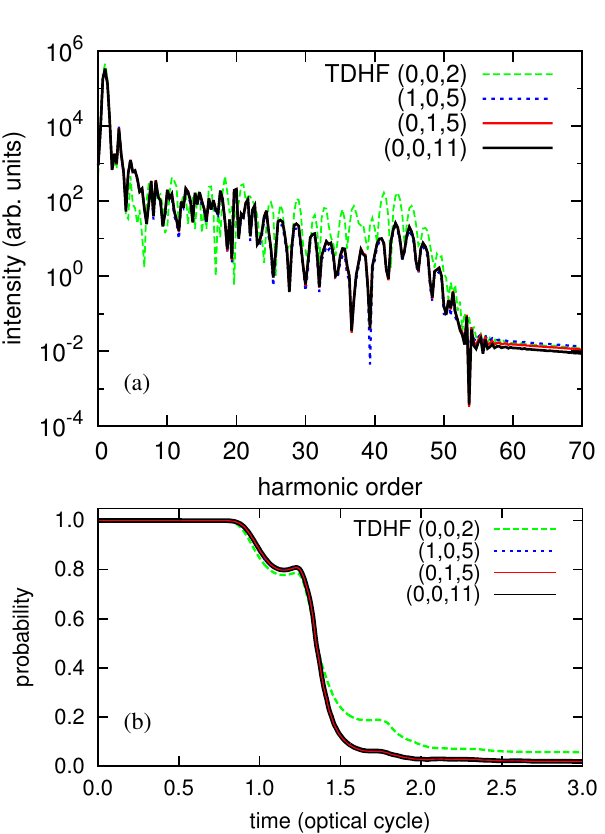}
\caption{\label{fig:be_hhg}(a) HHG spectra of Be exposed to an IR laser pulse
with a wavelength of 800 nm and an intensity of 3$\times$10$^{14}$ W/cm$^2$.
(b) Probability to find all electrons within a sphere of radius $R_0 = 20$ a.u.
Results of TDHF and TD-CASSCF with 
$(n_\textrm{fc}, n_\textrm{dc},n_\textrm{a})$={\color{black}$(0,0,2)$,} $(1,0,5)$, $(0,1,5)$, and $(0,0,11)$.
}%
\end{figure}

The HHG spectrum as well as the survival probability of beryllium (Fig.~\ref{fig:be_hhg}) for the same laser parameters as in Fig.~\ref{fig:be_tdhf} indicate rapid convergence of the TD-CASSCF method as a function of the size of the active space. We have checked that the employed spherical-FEDVR basis ($d_{\rm FE}=4, {\color{black} K_{\rm DVR}}=21,
R_{\rm box}=320$, and $L_{\rm max}=L_{\rm ee}=47$) and the time-step size (20000 steps per cycle) are
sufficient for convergence. 
Results for various active spaces $(n_\textrm{fc}, n_\textrm{dc},n_\textrm{a})=(1,0,5)$, $(0,1,5)$, and $(0,0,11)$
closely agree with each other indicating convergence with respect to correlation. By contrast, the TDHF method {\color{black}which lacks} correlation gives significant deviations. Unlike for helium where the TDHF underestimate{\color{black}s} HHG (Fig.~\ref{fig:he_w800_hhg_p0}) it overestimates HHG in beryllium (Fig.~\ref{fig:be_hhg}). This can be qualitatively easily understood as follows. In the strong-field regime and following the three-step model \cite{Lewenstein:1994} the harmonic yield $P(\omega)$ is expected to scale with the product of {\color{black}the} ionization rate at tunneling time $\sim|\dot{P}_0(t_i)|$ and the probability for finding the occupied ground state at recombination time $\sim P_0(t_r)$, 
\begin{eqnarray} \label{eq:hhg_yield}
P(\omega) \propto |\dot{P}_0(t_i)| P_0(t_r){\color{black}.}
\end{eqnarray}
In general, TDHF tends to underestimate the ionization rate
$|\dot{P}_0(t)|$ and therefore overestimate the ground-state probability
$\sim\!P_0(t)$ \cite{Kulander:1992}, as can be seen in Figs.~\ref{fig:he_w800_hhg_p0}~(c) and \ref{fig:be_hhg}~(b).
For the deeply bound helium electron (Fig.~\ref{fig:he_w800_hhg_p0}), $P_0(t) \approx 1$ holds for all $t$ for intensities $I {\color{black}\lesssim} 10^{14}$ W/cm$^2$. Thus, the ionization rate $|\dot{P}_0(t)|$ is the
dominant factor in {\color{black} Eq.~(\ref{eq:hhg_yield})}, leading to an underestimate of the harmonic intensity [Fig.~\ref{fig:he_w800_hhg_p0}~(c)].
%%% Then its underestimation by TDHF explains the
%%% underestimation of the harmonic intensity, 
On the other hand, for beryllium with a relatively weakly bound $2s$ valence electron, the atom experiences almost complete depletion of the
ground state $P_0$ at $t > 1.5T$, which controls now the efficiency of harmonic emission. Since
TDHF again largely overestimates $P_0$ especially at $t > 1.5T$
[Fig.~\ref{fig:be_hhg}~(b)] which is the relevant recombination time $t_r$ contributing to the plateau and cut-off region of the harmonic spectrum, it now overestimates the harmonic intensity. This example nicely illustrates the importance to go beyond the mean-field level of the TDHF approximation (or TDDFT where the same trends can be observed).

\subsection{Neon\label{subsec:numerical_ne}}
Finally{\color{black},} we calculate {\color{black}the} HHG spectrum of Ne subject to {\color{black}a} laser field with 
$\lambda = 800$ nm, $I_0 =$ 8.0$\times$10$^{14}$ W/cm$^2$, and $\tau =
3T$. We use {\color{black}the} spherical-FEDVR basis with
$d_{\rm FE}=4.0$, $K_{\rm DVR}=21$, and $L_{\rm max}=L_{\rm ee}=47$.
Simulations are performed in VG, and the HHG spectrum is calculated in
the acceleration form. For this 10-electron system we apply the TD-CASSCF method for {\color{black}various} active spaces with 
$(n_\textrm{fc}, n_\textrm{dc},n_\textrm{a})=(2,0,7)$, $(1,0,8)$, and $(0,0,9)$.
The 9 orbitals included are initially characterized as the atomic orbitals $1s$, $2s$, $2p_m$, $3s$, $3p_m$, with $m\in\{-1,0,1\}$. We consider 
either the $1s$ electrons to be frozen corresponding to $(1,0,8)$, or both $1s$ and $2s$ electrons to be frozen corresponding to $(2,0,7)$, or all electrons active $(0,0,9)$ (see Fig.~\ref{fig:cas}).
We obtain excellent agreement among all three active space configurations with a fixed total number of orbitals (Fig.~\ref{fig:ne}) indicating that the electron dynamics governing HHG is dominated by the {\color{black} $2p$} electrons. For a convergence check with respect to correlation, we also performed a TD-CASSCF calculation with a larger number of active orbitals with frozen $1s$ orbital $(1,0,13)$. We find near-perfect agreement among all TD-CASSCF calculations. Only the TDHF calculation shows systematic deviations underestimating the harmonic intensity as was the case for helium (see Fig.~\ref{fig:he_w800_hhg_p0}). 
The TD-CASSCF method is key to systematically explore the inclusion of correlation effects. The CI
dimension for the 
TD-CASSCF$(1,0,13)$ is about 1/8 ($\sim$ 500 thousands) of that for all
active MCTDHF method ($\sim$ 4 million) with the same number of occupied orbitals. 
In addition, the FC approximation brings further computational efficiency: 
freezing the deepest bound $1s$ orbital improves the stability of the propagation over the duration of the laser pulse.

\begin{figure}[!t]
\includegraphics[width=0.95\linewidth,clip]{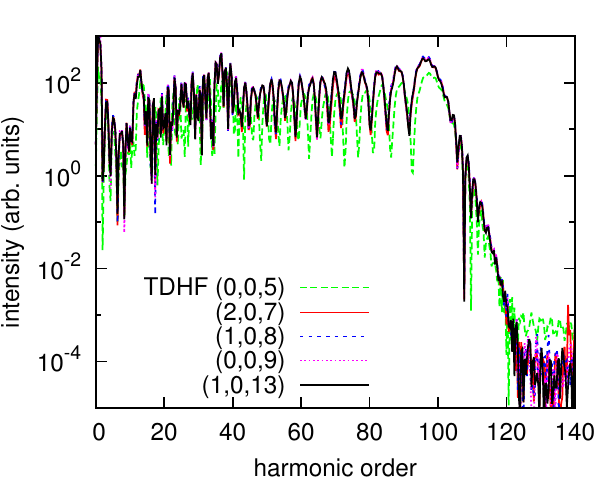}
\caption{\label{fig:ne}HHG spectrum of Ne driven by an IR laser pulse with $\lambda=$  800 nm, $I=$
8.0$\times$10$^{14}$ W/cm$^2$,
obtained with TDHF and TD-CASSCF with
$(n_\textrm{fc}, n_\textrm{dc},n_\textrm{a})$={\color{black}$(0,0,5)$,} $(2,0,7)$, $(1,0,8)$, $(0,0,9)$, and $(1,0,13)$.
Except for TDHF all results coincide within {\color{black}the} graphical resolution.
}%
\end{figure}

\section{Summary\label{sec:summary}}
We have presented an efficient 3D numerical implementation of the TD-CASSCF method for atoms subject to a
linearly polarized strong laser pulse. It features (i) a gauge-invariant frozen-core approximation allowing velocity gauge
simulations suitable for high-field phenomena, (ii) an efficient evaluation 
of the mean field operator with the computational cost scaling linearly with
the number of basis functions, and (iii) a split-operator method with a
stable implicit propagator for the one-electron Hamiltonian. We have also derived
and implemented the correction to the Ehrenfest expression {\color{black}the} of dipole velocity
and acceleration {\color{black}forms} to be used with frozen-core orbitals.

We have applied the present theory to He, Be, and Ne atoms, and obtained
survival probabilities and HHG spectra for intense NIR laser pulses,
convergent with respect to spatial and temporal 
discretization}, as well as the size of the active space. Our results indicate that the effect of correlation is relatively small for He, but
significant for Be and Ne, which underlines the general importance of methods
beyond the single-determinant TDHF approach. Correlation effects are
expected to be even more prominent for processes explicitly involving motions
of two or more electrons, such as nonsequential double ionization.
For accurately treating multi-electron and correlation effects, high levels of radial and angular resolution of the orbitals {\color{black}are} required. Otherwise such effects would be obscured by 
artifacts stemming from {\color{black}an} insufficient resolution of the orbitals as demonstrated in the
slow convergence of HHG spectrum of He with respect to the number of
spherical harmonics (Fig.~\ref{fig:he_hhg_w800_L1}).
{\color{black}
The present efficient implementation allows to
employ a large basis set (several thousands of radial grid points
$K_\textrm{rad}$ and a few hundred partial waves $L_{\rm max}$ are feasible on a single-node computer).
Further improvement could be achieved by employing multi-scale resolution grids recently implemented for molecules {\color{black}\cite{Sawada:2016}}.
%%%The final example of Ne actually demonstrates the first success of TD-CASSCF,
%%%which enables convergence assessment by restricting focus on important electrons. 
%%%The same example also shows a limitation of the
%%%active-full-CI based TD-CASSCF method that its application might be 
%%%up to 8 active electrons. 
The present gauge-invariant FC treatment and the efficient
propagation scheme can be straightforwardly applied to {\color{black}the} computationally even
less demanding TD-ORMAS method \cite{Sato:2015}. 
It will further extend the applicability to systems with a
larger number of electrons and/or broader range of laser parameters, and
will open the possibility of, e.g, {\color{black}\it ab initio} simulations of
XUV-NIR {\color{black} pump}-probe experiments.

\begin{acknowledgments}
We thank Othmar Koch for helpful discussions on propagation techniques {\color{black} and Johannes Feist for his work in the development of the TDSE helium code.}
This research is supported in part by Grant-in-Aid for Scientific
Research (No.~23750007, 23656043, 23104708, 25286064, 26390076, and
26600111) from the Ministry of Education, Culture, Sports, Science and
Technology (MEXT) of Japan, and also by {\color{black}the} Advanced Photon Science Alliance
(APSA) project commissioned by MEXT. This research is also partially
supported by the Center of Innovation Program from {\color{black}the} Japan Science and
Technology Agency, JST{\color{black}, and by CREST, JST}, and by the FWF-special research programs
F041-VICOM {\color{black}and} F049-NEXTlite, the FWF DK Solids4fun,
and the WWTF project MA14-002.
The TDSE results have been achieved using the Vienna Scientific Cluster (VSC).
\end{acknowledgments}

\appendix
\section{Second-quantization formulation\label{app:tdmcscf}}
We briefly review the formulation of the TD-MCSCF
\cite{Haxton:2015,Sato:2015} and TD-CASSCF theories \cite{Sato:2013}
within the framework of second quantization which allows for a compact
presentation of the theory. Using fermionic creation and annihilation
operators $\{\hat{a}_{\mu\sigma},\hat{a}^\dagger_{\mu\sigma}: \sigma \in \;\uparrow,\downarrow\}$
{\color{black} which change} the
occupation of spin orbitals $\{\psi_\mu\}\otimes\{\uparrow,
\downarrow\}$ with orthonormal spatial orbitals $\{\psi_\mu\}$, the
system Hamiltonian, Eq.~(\ref{eq:ham_1q}), can be rewritten as 
\begin{eqnarray}\label{eq:ham}
\hat{H} &=& \hat{h} + \hat{U} \\ &=&
{\color{black}
\sum_{\mu\nu} 
%(h^{\rm 1e})^\mu_\nu 
%h^\nu_\mu
h^\mu_\nu
\hat{E}^\mu_\nu + \frac{1}{2} \sum_{\mu\nu\gamma\lambda} 
%(V^{\rm 2e})^{\mu\gamma}_{\nu\lambda} 
%U^{\nu\lambda}_{\mu\gamma} 
U^{\mu\gamma}_{\nu\lambda}
\hat{E}^{\mu\gamma}_{\nu\lambda}}
,
\end{eqnarray}
where $h^\nu_\mu$ and $U^{\nu\lambda}_{\mu\gamma}$ are matrix elements
of the corresponding operators [see Eqs.~(\ref{eq:h1e_1q}) and (\ref{eq:v2_1q})]
%\begin{eqnarray}
%\label{eq:ham1e}
%%(h^{\rm 1e})^\mu_\nu 
%h^\nu_\mu
%= \int d\bm{r} \psi^*_\mu(\bm{r})
%h^{\rm 1e}\left(\bm{r}, \bm{p}\right)
%\psi_\nu(\bm{r}),
%\end{eqnarray}
%\begin{eqnarray}
%\label{eq:ham2e}
%%(V^{\rm 2e})^{\mu\gamma}_{\nu\lambda} 
%V^{\rm 2e}_{\mu\nu\gamma\lambda}
%= \int\!\!\!\int \! d\bm{r}_1 d\bm{r}_2
%\frac{
%\psi^*_\mu(\bm{r}_1) \psi_\nu(\bm{r}_1)
%\psi^*_\gamma(\bm{r}_2) \psi_\lambda(\bm{r}_2)
%}{|\bm{r}_1 - \bm{r}_2|},
%\end{eqnarray}
{\color{black},}
$
\hat{E}^\mu_\nu = \sum_\sigma
\hat{a}^\dagger_{\mu\sigma}\hat{a}_{\nu\sigma}
$, and
$
\hat{E}^{\mu\gamma}_{\nu\lambda} = \sum_{\sigma\sigma^\prime}
\hat{a}^\dagger_{\mu\sigma}\hat{a}^\dagger_{\gamma\sigma^\prime}
\hat{a}_{\lambda\sigma^\prime}\hat{a}_{\nu\sigma}
$.
%%% Throughout this paper, operators and wave functions in the real space are
%%% denoted by $H,h^{\rm 1e}(\bm{r},\bm{p})$ and $\Psi,\psi_\mu(\bm{r})$,
%%% etc, while those in the second quantization by $\hat{H},\hat{h}^{\rm
%%% 1e}$ and $|\Psi\rangle,|\psi_\mu\rangle$.
%%% $N_b$ is identical to
%%% the number of basis functions (or grid points) to expand orbitals.
%%% The convergence of results with respect to $N_b$, or equivalently, the spatial
%%% resolution, has to be assessed carefully as will be discussed in Sec.~\ref{sec:numerical}.
%%% In principle, $N_b$ should be infinity to have exact equivalence of
%%% first- and second-quantized Hamiltonian, Eq.~(\ref{eq:ham_1q}) and
%%% (\ref{eq:ham}). In practice, however, $N_b$ is the number of basis
%%% functions to expand orbitals, and the convergence of results with
%%% respect to $N_b$, the spatial resolution, has to be assessed carefully
%%% as will be discussed in Sec.~\ref{sec:numerical}.
%%% The full set of orbitals $\{\psi_\mu\}$ are classified into $n$ occupied orbitals $\{\psi_p;
%%% p=1,2,\cdot\cdot\cdot,n\}$, and remaining $N_b-n$
%%% virtual orbitals $\{\psi_a; a=n+1,n+2,\cdot\cdot\cdot,N_b\}$, and
%%% the determinant $\Phi_I$ of Eq.~(\ref{eq:mcscf}) is built
%%% from the occupied orbitals only. 
%%% Then the EOMs for the general TD-MCSCF method is derived based on the
%%% time-dependent variational principle.

In the TD-MCSCF method, the full set of orbitals
$\{\psi_\mu\}$ is classified as $n$ occupied orbitals $\{\psi_p;
p=1,2,\ldots,n\}$ and the remaining as 
virtual orbitals $\{\psi_a; a=n+1,n+2,\ldots\}$.
The determinant $\Phi_I$ of Eq.~(\ref{eq:mcscf}) is built
from the occupied orbitals only,
{\color{black}
\begin{eqnarray}\label{eq:mcscf_2q}
|\Psi\rangle = \sum_I {\color{black}|\Phi_I\rangle} C_I,
\end{eqnarray}
\begin{eqnarray}\label{eq:det}
\vert \Phi_I \rangle &=& %%% \hat{I}
\prod_\sigma\prod_p
(\hat{a}^\dagger_{p\sigma})^{I_{p\sigma}} \vert 0\rangle,
\end{eqnarray}
where $| 0 \rangle$ represents the vacuum state,
$I_{p\sigma} = \{0,1\}$, and $\sum_\sigma\sum_p
I_{p\sigma} = N$.
}
%The essence of the MCSCF method is the variational separation of occupied and
%virtual orbital spaces; the CI problem is solved within the optimized
%occupied orbital space.
%%\delta S = 0; \hspace{.5em}
%\delta\int\!dt \langle\Psi| \hat{H} - {\rm  i}\frac{\partial}{\partial
%t} |\Psi\rangle = 0.
%\end{eqnarray}
\\

Time derivatives of orbitals are conveniently represented by the Hermitian
one-electron operator 
$\hat{X}$
\cite{Miranda:2011a, Sato:2013}
\begin{subequations}\label{eqs:td_general}
\begin{eqnarray}\label{eq:tdmo_general}
i|\dot{\psi}_p\rangle = \hat{X}|\psi_p\rangle, \hspace{.5em}
{\color{black}
\hat{X}=\sum_{\mu\nu} X^\mu_\nu \hat{E}^\mu_\nu.
}
\end{eqnarray}
%\begin{eqnarray}\label{eq:xop}
%\hat{X} = \sum_{\sigma\mu\nu} X_{\mu\nu} %%% \hat{E}_{\mu\nu}
%\hat{a}^\dagger_{\mu\sigma}\hat{a}_{\nu\sigma}, \hspace{.5em}
%X_{\mu\nu} = i\langle\psi_\mu|\dot{\psi}_\nu\rangle.
%\end{eqnarray}
The EOMs for the TD-MCSCF method are derived based on the
time-dependent variational principle, i.e., by requiring the 
action integral of Eq.~(\ref{eq:action})
to be stationary with respect to the variation of CI coefficients and orbitals,
$\partial S/\partial C_I = 0$, $\partial S/\partial \Delta_{\mu\nu}=0$, where
$\Delta_{\mu\nu}$ is an anti-Hermitian matrix generating
orthonormality-conserving variations of orbitals as $\delta\psi_p =
\sum_\mu \psi_\mu \Delta_{\mu p}$ \cite{Sato:2013,Sato:2015}.
The resulting EOM for the CI coefficients is
\begin{eqnarray}\label{eq:tdci_general}
{\color{black}
i\dot{C}_I = \langle \Phi_I|\hat{H} - \hat{X}|\Psi\rangle,
}
\end{eqnarray}
and the EOMs for the orbitals Eq.~(\ref{eq:tdmo_general}) are determined by
%\begin{subequations}\label{eqs:td_general}
%\begin{eqnarray}\label{eq:tdmo_general}
%i|\dot{\psi}_p\rangle = \hat{X}|\psi_p\rangle,
%\end{eqnarray}
%\begin{eqnarray}\label{eq:tdci_general}
%i\dot{C}_I = \langle \Phi_I|\hat{H} + \hat{X}|\Psi\rangle,
%\end{eqnarray}
\begin{eqnarray}\label{eq:tdorb_general1}
\sum_{\gamma\lambda} A^{\nu\gamma}_{\mu\lambda} X^{\lambda}_{\gamma} = B^\nu_\mu,
\end{eqnarray}
\end{subequations}
where
\begin{eqnarray}\label{eq:tdorb_general1_coeff}
A^{\nu\gamma}_{\mu\lambda} &=&
\langle\Psi^\nu_\mu|(1-\hat{\Pi}) |\Psi^\gamma_\lambda\rangle -% \nonumber \\ &-&
\langle\Psi^\lambda_\gamma|(1-\hat{\Pi}) |\Psi^\mu_\nu\rangle,
\end{eqnarray}
\begin{eqnarray}\label{eq:tdorb_general1_rhs}
B^\nu_\mu = 
\langle\Psi^\nu_\mu| (1-\hat{\Pi}) \hat{H} |\Psi\rangle -
\langle\Psi| \hat{H} (1-\hat{\Pi}) |\Psi^\mu_\nu\rangle,
\end{eqnarray}
%\begin{eqnarray}\label{eq:dpsidrot}
%|\Psi^\mu_\nu\rangle = %%% \hat{E}_{\mu\nu}
%\sum_\sigma
%\hat{a}^\dagger_{\mu\sigma}\hat{a}_{\nu\sigma}
%|\Psi\rangle,
%\end{eqnarray}
with $|\Psi^\mu_\nu\rangle=\hat{E}^\mu_\nu|\Psi\rangle$ and $\hat{\Pi}=\sum_I|\Phi_I\rangle\langle \Phi_I|$ is the projector onto the
space spanned by those Slater determinants included in Eq.~(\ref{eq:mcscf_2q}). Following
Eqs.~(\ref{eq:tdmo_general}) and (\ref{eq:tdci_general}) the total time derivative of the wave function is compactly 
written as
\begin{eqnarray}\label{eq:tdtot_mcscf}
i|\dot{\Psi}\rangle %= \sum_I^\Pi|\Phi_I\rangle\dot{C}_I + \hat{X}|\Psi\rangle %\nonumber \\
= \hat{\Pi}\hat{H}|\Psi\rangle + (1-\hat{\Pi})\hat{X}|\Psi\rangle.
\end{eqnarray}
This expression shows that the time derivative of the wave function can
be separated into two orthogonal contributions. The first term in
Eq.~(\ref{eq:tdtot_mcscf}) gives the time evolution of the wave function
in the subspace $\hat{\Pi}$ spanned by the Slater determinants $\vert \Phi_I\rangle$.
Due to the time-dependence of the orbitals this subspace
itself is time-dependent giving rise to the second term in
Eq.~(\ref{eq:tdtot_mcscf}). This additional contribution directly
reflects the strength of the TD-MCSCF method compared to the approach
with time-independent orbitals where the evolution of the wave function
is confined solely within the initial subspace.  
Equation (\ref{eq:tdtot_mcscf}) shows that the component of
$\hat{X}\vert \Psi \rangle$ that lies within $\hat{\Pi}$ cannot
contribute to the evolution of the wave function. This gives rise to the
distinction {\color{black}between {\it redundant} and {\it
nonredundant}} orbital pairs $\{\mu,\nu\}$. Only if $(1-\hat{\Pi})\hat{E}^\mu_\nu|\Psi\rangle\ne0$
or $(1-\hat{\Pi})\hat{E}^\nu_\mu|\Psi\rangle\ne0$, 
the matrix element $X_\nu^\mu$ influences the evolution of the wave
function [see Eq.~(\ref{eq:tdtot_mcscf})] and is, therefore, called {\it
nonredundant}. Otherwise it is called {\it redundant}. For redundant
pairs {\color{black}$\{\mu,\nu\}$}, as is clear from Eqs.~(\ref{eq:tdorb_general1_coeff}) and
(\ref{eq:tdorb_general1_rhs}), both $A^{\nu\gamma}_{\mu\lambda}$ and
$B^\nu_\mu$ vanish, 
thus Eq.~(\ref{eq:tdorb_general1}) [or Eq.~(\ref{eq:tdorb_general2}) see below] reduces
to a trivial identity, and correspondingly $X^\nu_\mu$ may be an
arbitrary Hermitian matrix element. Therefore, Eq.~(\ref{eq:tdorb_general1}) [or
Eq.~(\ref{eq:tdorb_general2})] should be solved only for nonredundant pairs
{\color{black}$\{\mu,\nu\}$}.

%One can see, from Eq.~(\ref{eq:tdtot_mcscf}), that only the nonredundant
%part of $\hat{X}$ contributes to the total time derivative of the wave
%function. 
%%% See Ref.~\cite{Sato:2015} for in-depth analysis of the
%%% concept of the (non-)redundancy and its optimal use in constructing MCSCF wave functions. 
%}
%{\color{black}
\section{Equivalent forms of EOMs\label{app:oorot}}
For the development of {\color{black}the} split operator method in
Sec.~\ref{subsec:split}, we also present an equivalent but different
form of EOMs. 
%%% Inserting the second-quantized expression of
%%% $\hat{h}^{1e}$ into Eq.~(\ref{eq:tdorb_general1_rhs}) and using
%%% Eqs.~(\ref{eq:tdorb_general1_coeff}) and (\ref{eq:dpsidrot}), one
%%% readily obtains
First we note
\begin{eqnarray} \label{eq:bmat2e}
B^\nu_\mu &=&
%\sum_{\gamma\lambda}\langle\Psi_{\nu\mu}| (1-\hat{\Pi})
%|\Psi_{\gamma\lambda}\rangle h^{\rm 1e}_{\gamma\lambda} \nonumber \\ &-&
%\sum_{\gamma\lambda}h^{\rm 1e}_{\gamma\lambda} \langle\Psi_{\lambda\gamma}| (1-\hat{\Pi})
%|\Psi_{\mu\nu}\rangle \nonumber \\ &+&
%\langle\Psi_{\nu\mu}| (1-\hat{\Pi}) \hat{V}^{\rm 2e} |\Psi\rangle -
%\langle\Psi| \hat{V}^{\rm 2e} (1-\hat{\Pi}) |\Psi_{\mu\nu}\rangle \nonumber \\ &=&
\sum_{\gamma\lambda} A^{\nu\gamma}_{\mu\lambda} h^\lambda_\gamma 
+ \tilde{B}^\nu_\mu,
\end{eqnarray}
where $\tilde{B}$ is given by Eq.~(\ref{eq:tdorb_general1_rhs}) with
$\hat{H}$ replaced with $\hat{U}$. Inserting this into
Eq.~(\ref{eq:tdorb_general1}) and introducing an auxiliary operator
\begin{eqnarray}\label{eq:yop}
\hat{R} = \hat{X} {\color{black}-} \hat{h}, 
\end{eqnarray}
%we can rewrite the EOMs of the orbitals and the CI-coefficients as
we can rewrite Eqs.~(\ref{eqs:td_general}) as
%%% the general EOMs are written as
\begin{subequations}\label{eqs:td_general_split}
\begin{eqnarray}\label{eq:tdmo_general_split}
i|\dot{\psi}_p\rangle = \hat{h}|\psi_p\rangle + \hat{R}|\psi_p\rangle,
\end{eqnarray}
\begin{eqnarray}\label{eq:tdci_general_split}
i\dot{C}_I = \langle \Phi_I|\hat{U} {\color{black}-} \hat{R}|\Psi\rangle,
\end{eqnarray}
where $\hat{R}$ is determined by 
\begin{eqnarray}\label{eq:tdorb_general2}
\sum_{\gamma\lambda} A^{\nu\gamma}_{\mu\lambda} R^{\lambda}_{\gamma} = \tilde{B}^\nu_\mu.
%\sum_{\gamma\lambda} A_{\mu\nu,\gamma\lambda} X_{\gamma\lambda} = -iB_{\mu\nu},
\end{eqnarray}
\end{subequations}
The transformed EOMs~(\ref{eqs:td_general_split}) are the generalization
of those used in the variational splitting method for MCTDHF
\cite{Lubich:2004,Caillat:2005,Koch:2010}.
Yet another form of the EOMs can be derived by splitting off only the
time-independent atomic Hamiltonian $\hat{h}_0$ instead of $\hat{h}$. We have implemented various choices for the 
EOMs and the redundant matrix elements $X^\mu_\nu$,
$R^\mu_\nu$. Among them, the second form {\color{black}Eq.}~(\ref{eqs:td_general_split}){\color{black},} with the split-operator method described in
Sec.~\ref{subsec:split}{\color{black},} was found generally most robust. In this case, all stiff
derivative operators (kinetic energy and laser-electron
interaction in VG) and the singular nucleus-electron
interaction are treated with the stable implicit method.
%%% We note, however, that Eqs.~(\ref{eqs:td_general_split}) are just a
%%% rearrange of Eqs.~(\ref{eqs:td_general}) using Eq.~(\ref{eq:bmat2e}); 
The EOMs {\color{black}Eq.}~(\ref{eqs:eom_split}) for the TD-CASSCF method are obtained by
applying the general TD-MCSCF EOMs~(\ref{eqs:td_general_split}) to the
following CASSCF ansatz for the total wave function, 
\begin{eqnarray}\label{eq:casscf_2q}
|\Psi_\textrm{CAS}\rangle = \hat{\Phi}_\textrm{fc} \hat{\Phi}_\textrm{dc} \sum_I |\Phi_I\rangle C_I,
\end{eqnarray}
where $\hat{\Phi}_\textrm{fc}\equiv\prod_i^{\rm
FC}\hat{a}^\dagger_{i\uparrow}\hat{a}^\dagger_{i\downarrow}$,
$\hat{\Phi}_\textrm{dc}\equiv\prod_i^{\rm
DC}\hat{a}^\dagger_{i\uparrow}\hat{a}^\dagger_{i\downarrow}$, and
$\vert \Phi_I\rangle = \prod_\sigma\prod_t
(\hat{a}^\dagger_{t\sigma})^{I_{t\sigma}} \vert 0\rangle$, with
$\sum_{t\sigma} I_{t\sigma}=N_\textrm{a}$.
There still remains the freedom to choose {\color{black}the} redundant part of
$\hat{R}$, which is {\color{black}set to zero} in Eqs.~(\ref{eqs:eom_split}).

\section{Conservation of the orbital magnetic quantum number\label{app:m_const}}
We require that, {\it at any given time t}, both  the total wave function
$|\Psi\rangle$ and each spatial orbital $|\psi_\mu\rangle$, are
eigenfunctions of the component of the orbital angular momentum parallel
to the laser polarization direction ($z$-axis),
\begin{eqnarray}
\hat{l}_z=-i\sum_{\mu\nu} \hat{E}^\mu_\nu
%\hat{a}^\dagger_{\mu\sigma}\hat{a}_{\nu\sigma}
\int d\bm{r} \psi^*_\mu(\bm{r}) \frac{\partial}{\partial \phi} \psi_\nu(\bm{r}),
%\langle\psi_\mu|\frac{\partial}{\partial\phi}|\psi_\nu\rangle 
%\hat{E}^\mu_\nu,
\end{eqnarray}
with eigenvalues $M_{\rm tot}$ and $\{m_\mu\}$, respectively.
%\begin{eqnarray}
%\hat{l}_z|\psi_\mu\rangle = m_\mu |\psi_\mu\rangle, 
%%-i\frac{\partial}{\partial \phi}\psi_\mu(\bm{r}) = m_\mu
%% \psi_\mu(\bm{r}), 
%\end{eqnarray}
%\begin{eqnarray}
%\hat{l}_z |\Psi\rangle = M_{\rm tot} |\Psi\rangle,
%\end{eqnarray}
%where $\hat{l}_z=-i\sum_\mu\langle\psi_\mu|\partial/\partial\phi|\psi_\mu\rangle
%\hat{E}^\mu_\mu$.
We show that within the framework of general
TD-MCSCF methods and for interactions {\color{black} which are} rotationally symmetric about the
$z$-axis ({\color{black}e.g,} laser-matter interaction with linear polarization along the $z$-axis) $M_{\rm tot}$ as well as $m_\mu$ are constant for all times.
We note that the total Hamiltonian $\hat{H}$ commutes with
$\hat{l}_z$, and any two eigenstates of $\hat{l}_z$, $(|\Psi^\prime\rangle,
|\Psi^{\prime\prime}\rangle)$ with different eigenvalues $M^\prime,
M^{\prime\prime}$ are orthogonal to each other. {\color{black}Thus}, $\hat{l}_z$ also
commutes with the CI-space projector $\hat{\Pi}$.
Now we consider Eq.~(\ref{eq:tdorb_general1}), which is solved for 
nonredundant pairs {\color{black}$\{\mu,\nu\}$} as
\begin{eqnarray}\label{eq:tdorb_general1_solved}
X^\nu_\mu &=& \sum_{\gamma\lambda} {(A^{-1})}^{\nu\gamma}_{\mu\lambda} B^\lambda_\gamma,
%\sum_{\gamma\lambda} A^{\nu\gamma}_{\mu\lambda} X^{\lambda}_{\gamma} = B^\nu_\mu,
\end{eqnarray}
where $A^{-1}$ and $B$ are regarded as a matrix and a vector, respectively.
The element $B^\lambda_\gamma$ {\color{black}is nonzero only if $m_\gamma = m_\lambda$,}
since otherwise the state $|\Psi^\gamma_\lambda\rangle=\hat{E}^\gamma_\lambda|\Psi\rangle$ would have an eigenvalue
different from $M_{\rm tot}$,
\begin{eqnarray}
\hat{l}_z|\Psi^\gamma_\lambda\rangle = (M_{\rm tot} + m_\gamma - m_\lambda) |\Psi^\gamma_\lambda\rangle.
\end{eqnarray}
A similar argument applied to Eq.~(\ref{eq:tdorb_general1_coeff}) yields
that 
{\color{black}
%the matrix $A$ is block diagonal for the part where first
%and second unified indices satisfy ($m_\mu$$=$$m_\nu$, $m_\lambda$$=$$m_\gamma$)
%and the part ($m_\mu$$\neq$$m_\nu$, $m_\lambda$$\neq$$m_\gamma$).
the element $A^{\nu\gamma}_{\mu\lambda}$, and thus
$(A^{-1})^{\nu\gamma}_{\mu\lambda}$, is nonzero only if $m_\mu - m_\nu
=m_\lambda - m_\gamma$. 
}
Consequently, $X^\nu_\mu$ {\color{black}[Eq.~(\ref{eq:tdorb_general1_solved})]} vanishes if {\color{black}$m_\mu \neq m_\nu$}. As a result, 
orbitals propagated according to Eq.~(\ref{eq:tdmo_general})
conserve $\{m_\mu\}$. Now, it follows from Eq.~(\ref{eq:tdtot_mcscf}) that
$\hat{l}_z\dot{|\Psi\rangle} = M_{\rm tot}\dot{|\Psi\rangle}$, since $[\hat{l}_z,
\hat{X}]=0$. Therefore, the total projection $M_{\rm tot}$ is also
conserved.

\section{Ehrenfest theorem for the TD-MCSCF method\label{app:ehrenfest}}
We consider the time derivative $d\langle\hat{O}\rangle/dt=d
\langle\Psi|\hat{O}|\Psi\rangle /dt$ of the expectation value of an
operator $\hat{O}$,
\begin{eqnarray}\label{eq:expo}
\frac{d}{dt}\langle \hat{O} \rangle =
\langle\dot{\Psi}|\hat{O}|\Psi\rangle +
\langle\Psi|\hat{O}|\dot{\Psi}\rangle +
\langle\Psi|\frac{\partial\hat{O}}{\partial t}|\Psi\rangle. 
\end{eqnarray}
For {\color{black}{an}} exact solution of {\color{black}{the}} TDSE, inserting
$i|\dot{\Psi}\rangle=\hat{H}|\Psi\rangle$ and its hermitian conjugate
into Eq.~(\ref{eq:expo}), one can derive the Ehrenfest theorem, 
\begin{eqnarray}\label{eq:ehrenfest}
i\frac{d}{dt}\langle \hat{O} \rangle &=& 
\langle\Psi|[\hat{O}, \hat{H}]|\Psi\rangle +
i\langle\Psi|\frac{\partial\hat{O}}{\partial t}|\Psi\rangle.
\end{eqnarray}
%%% In particular for $\hat{O} = \hat{z}$, the canonical commutation relation gives
%%% \begin{eqnarray}\label{eq:ehrenfest_vel}
%%% \frac{d}{dt}\langle \hat{z} \rangle = \left\{
%%% \begin{array}{cc}
%%% \langle\Psi|\hat{p}_z|\Psi\rangle & ({\rm LG})\\
%%% \langle\Psi|\hat{p}_z|\Psi\rangle + N A(t) & ({\rm VG})
%%% \end{array}
%%% \right.,
%%% \end{eqnarray}
%%% \begin{eqnarray}\label{eq:ehrenfest_acc}
%%% \frac{d^2}{dt^2}\langle \hat{z} \rangle = -
%%% \langle\Psi|\frac{\partial \hat{V}_0}{\partial z}|\Psi\rangle - N E(t).
%%% \end{eqnarray}

For the TD-MCSCF method, inserting
Eq.~(\ref{eq:tdtot_mcscf}) and its hermitian conjugate into
Eq.~(\ref{eq:expo}), one obtains 
\begin{eqnarray}\label{eq:ehrenfest_mcscf}
\label{eq:ehrenfest_mcscf_ci}
i\frac{d}{dt}\langle \hat{O} \rangle &=& 
\label{eq:ehrenfest_mcscf_line1}
\langle\Psi|\hat{O}\hat{\Pi}\hat{H} - \hat{H}\hat{\Pi}\hat{O}
|\Psi\rangle \nonumber \\ &+&
\langle\Psi|\hat{O}(1-\hat{\Pi})\hat{X} - \hat{X}(1-\hat{\Pi})\hat{O}
|\Psi\rangle + i\langle\Psi|\frac{\partial
\hat{O}}{\partial t}|\Psi\rangle \nonumber \\ &=&  
\label{eq:ehrenfest_mcscf_line2}
\langle\Psi|[\hat{O}, \hat{H}]|\Psi\rangle + 
i\langle\Psi|\frac{\partial \hat{O}}{\partial t}|\Psi\rangle \nonumber \\
 &+&
\sum_{\mu\nu}\left(
\sum_{\gamma\lambda} A^{\nu\gamma}_{\mu\lambda} X^{\lambda}_{\gamma} - B^\nu_\mu
\right) O_{\mu\nu},  \nonumber \\ &=&
\label{eq:ehrenfest_mcscf_line3}
\langle\Psi|[\hat{O}, \hat{H}]|\Psi\rangle + 
i\langle\Psi|\frac{\partial \hat{O}}{\partial t}|\Psi\rangle
+ i\Delta(\dot{O}),
\end{eqnarray}
where the orbital EOMs of TD-MCSCF, Eqs.~(\ref{eq:tdorb_general1})-(\ref{eq:tdorb_general1_rhs}){\color{black},}
are used for the second equality.
The third line defines {\color{black}the} quantity $\Delta(\dot{O})$, representing the difference
from the Ehrenfest expression~(\ref{eq:ehrenfest}), 
\begin{eqnarray}\label{eq:gbf}
\Delta(\dot{O}) &=&
-i\sum_{\mu\nu}\left(
\sum_{\gamma\lambda} A^{\nu\gamma}_{\mu\lambda} X^{\lambda}_{\gamma} - B^\nu_\mu
\right) O^\mu_\nu.
\end{eqnarray}
In the absence of frozen-core orbitals, Eq.~(\ref{eq:tdorb_general1}) holds for
all nonredundant pairs {\color{black}$\{\mu,\nu\}$}, thus $\Delta(\dot{O}) = 0$, and
Eq.~(\ref{eq:ehrenfest_mcscf}) reduces to Eq.~(\ref{eq:ehrenfest}). 
This establishes the applicability of the Ehrenfest theorem to TD-MCSCF wave functions.
With frozen-core orbitals $\{\psi_k\}$, however, 
the equality (\ref{eq:tdorb_general1}) does not hold for pairs
{\color{black}$\{\mu, k\}$} and {\color{black}$\{k, \mu\}$}, and thus $\Delta(\dot{O}) \neq 0$. 
Instead we have
\begin{eqnarray}\label{eq:gbf_fc}
\Delta(\dot{O}) &=& 
i\sum_{k\mu}O^k_\mu
\langle\Psi|[\hat{E}^\mu_k, \hat{H}-\hat{X}]|\Psi\rangle \nonumber \\ &+& 
i\sum_{k\mu}O^\mu_k 
\langle\Psi|[\hat{E}^k_\mu , \hat{H}-\hat{X}]|\Psi\rangle \nonumber \\ &=&
2i \sum_k \langle\psi_k|\left[\hat{O},\hat{h}+\hat{F}-\hat{X}\right]|\psi_k\rangle
-2{\rm Im} \sum_{kp} O^k_p B^p_k, \nonumber \\
\end{eqnarray}
where $B^p_k=\sum_q(h^q_k+F^q_k-X^q_k) D^p_q$.
%{\color{black}
%$\hat{X}=0$ for LG, and $\hat{X}=E(t)\hat{z}$ for VG.}
%%% $\hat{F}_k = \hat{h}^{\rm 1e}+\hat{X}+\hat{V}_k$,
%%% $F_{ki} = 2(h^{\rm 1e}_{ki}+X_{ki}+V_{ki})$, 
%%% and $F_{kt} = \sum_{u} (h^{\rm 1e}_{ku}+X_{ku}+V_{ku}) D_{ut}$.
%Ingredients of Eq.~(\ref{eq:gbf_fc}) are necessary for propagating
%orbitals, thus can be evaluated with little additional cost.
%
For a simple physical interpretation of $\Delta$, we consider the LG and
temporarily neglect the indistinguishability of core and active electrons. In this case 
Eq.~(\ref{eq:gbf_fc}) reduces to
\begin{eqnarray}\label{eq:gbf_fc_approx}
\Delta(\dot{O}) &\approx&
i\langle\Phi_\textrm{fc}|[\hat{O}, \hat{h} + \hat{V}_\textrm{a}]|\Phi_\textrm{fc}\rangle,
\end{eqnarray}
where $\Phi_\textrm{fc}$ is the {\color{black}FC} part of the wave function, and
$\hat{V}_\textrm{a}$ is the electrostatic
potential of active electrons, given in real space as
\begin{eqnarray}\label{eq:coulomb_act}
%V_\textrm{c}(\bm{r}) &=& \int d\bm{r}^\prime \frac{\rho_\textrm{c}(\bm{r}^\prime)}{|\bm{r}-\bm{r}^\prime|}, \\
V_\textrm{a}(\bm{r}) &=& \int d\bm{r}^\prime \frac{\rho_\textrm{a}(\bm{r}^\prime)}{|\bm{r}-\bm{r}^\prime|},
\end{eqnarray}
with $\rho_\textrm{a}$ being the density of active electrons.
Within this approximation, we obtain,
\begin{eqnarray}\label{eq:gbf_fc_approx_z}
\Delta(\dot{z}) \approx \langle\Phi_\textrm{fc}|\frac{\partial \color{black}\hat{h}}{\partial p_z}|\Phi_\textrm{fc}\rangle
= \langle\Phi_\textrm{fc}|\hat{v}|\Phi_\textrm{fc}\rangle,
\end{eqnarray}
\begin{eqnarray}\label{eq:gbf_fc_approx_p}
\Delta(\dot{p}_z) &\approx& 
 \langle\Phi_\textrm{fc}|
 \frac{\partial \hat{V}_0}{\partial z} +
 \frac{\partial \hat{V}_{\rm ext}}{\partial z} +
 \frac{\partial \hat{V}_\textrm{a}}{\partial z}
 |\Phi_\textrm{fc}\rangle.
\end{eqnarray}
{\color{black}The FC expectation value of the kinematic momentum operator $\hat{v}$ [Eq.~(\ref{eq:gbf_fc_approx_z})]} 
is time independent in general and
vanishes for atomic systems, i.e., $\Delta(\dot{z})\approx 0$.
Equation~(\ref{eq:gbf_fc_approx_p}) can be interpreted as the binding
force $f_{\rm b}$ discussed in Sec.~\ref{subsec:numerical_be}, 
\begin{eqnarray} \label{eq:force_fc_again}
f_{\rm b} = -f_{\rm nc} - f_{\rm lc} - f_{\rm ac}.
\end{eqnarray}

\section{Matrix elements in the spherical-FEDVR basis\label{app:fedvr}}
Nonzero matrix elements of one-electron operators in the spherical-FEDVR
basis read
\begin{eqnarray}\label{eq:h0_fedvr}
 (\bm{h}_0)^{klm}_{k^\prime l m} &=&
-\frac{1}{2}\nabla^2_{kk^\prime} + \delta_{kk^\prime}
\left\{\frac{l(l+1)}{2r^2_k} - \frac{Z}{r_k}\right\},
\end{eqnarray}
\begin{eqnarray}\label{eq:vext_length_fedvr}
(\bm{V}_{\rm ext}^{\rm LG})^{klm}_{kl^\prime m} &=&
%(\bm{V}_{\rm ext}^{\rm LG})_{klm}^{k,l+1,m} =
r_k E(t) \alpha_{lm},
\end{eqnarray}
%%%
\begin{eqnarray}\label{eq:vext_velocity_fedvr}
&&(\bm{V}_{\rm ext}^{\rm VG})^{klm}_{k^\prime l^\prime m} =
%(\bm{V}_{\rm ext}^{\rm VG})_{klm}^{k^\prime, l+1, m} \nonumber \\ && =
-iA(t) \left\{\nabla_{kk^\prime} - \delta_{kk^\prime}\frac{(l+1)}{r_k}\right\}\alpha_{lm},
\end{eqnarray}
\begin{eqnarray}
\alpha_{lm} = \sqrt{\frac{(l+1)^2 - m^2}{(2l+1)(2l+3)}},
\end{eqnarray}
where $l^\prime = l + 1$ and $(\bm{V}_{\rm ext})^{k^\prime l^\prime
m}_{klm} = (\bm{V}_{\rm ext})_{k^\prime l^\prime m}^{klm*}$.
See, {\color{black} e.g.}, Ref.~\cite{Schneider:2011} for FEDVR matrix elements of
derivative operators, 
$\nabla_{kk^\prime} = \langle f_k|\nabla_r| f_{k^\prime}\rangle$ and
$\nabla^2_{kk^\prime} = \langle f_k|\nabla^2_r|f_{k^\prime}\rangle$. 
\\

The operator $W^p_q(r_k,\theta_j)$ (see Eq.~\ref{eq:meanfield}) describing the electron-electron interaction is evaluated as follows.
%The two electron part $\bm{F}^{\rm 2e}_p\bm{\varphi}_p$ of Eq.~(\ref{eq:eq:eom_matrix})
%is evaluated as follows. 
%old by (i) first solving Poisson's equation, Eq.~(\ref{eq:poisson})
%old and (ii) then contracting them with RDMs through
%old Eqs.~(\ref{eqs:fock2}) and (\ref{eq:core_fock2}). 
%old While Poisson's equation [step (i)] can be efficiently solved 
%old in the spherical-FEDVR basis of Eq.~(\ref{eq:fedvr}) \cite{McCurdy:2004,Hochstuhl:2011},
%old the step (ii) is most naturally performed in the real space.
%old Based on the above observation, 
First we transform orbitals into the two-dimensional
$(r_k,\theta_j)$ grid representation,
\begin{eqnarray}
\varphi_p(r_k,\theta_j) = \frac{1}{\sqrt{w^\textrm{rad}_k}}\sum_{l=|m_p|}^{L_{\rm max}} G_{j,lm_p} \bm{\varphi}^{kl}_p,
\end{eqnarray}
where $G_{j,lm} = P_{lm}(\cos\theta_j)$, $P_{lm}$ is an associated
Legendre polynomial, and $\{\cos\theta_j\}$ and $\{w^{\rm ang}_j\}$
(appearing below) are {\color{black}the} nodes and weights of the Gauss-Legendre
quadrature, respectively, of order {\color{black}$L_{\rm ee}$}.
Note that the original three-dimensional orbital is given by $\psi_p(r,\theta,\phi)
= e^{im_p\phi}\varphi_p(r,\theta)/r$. 
Next, the pair densities $\{{\color{black}\rho^p_q} \equiv \varphi^*_p \varphi_q\}$ are obtained by
multiplications on the grid, and transformed back into the ${kl}$ basis,
\begin{eqnarray}
(\bm{\rho}^p_q)_{kl} = \sum_{j=0}^{L_{\rm ee}} G^{-1}_{lM^p_q,j} \rho^p_q(r_k,\theta_j),
\end{eqnarray}
where $G^{-1}_{lm,j} = w^{\rm ang}_j P^*_{lm}(\cos\theta_j)$, and
$M^p_q = -m_p + m_q$. The transformed pair densities serve as
the source for the radial Poisson equation
\cite{McCurdy:2004,Hochstuhl:2011} for each partial
wave $lM^p_q$,
\begin{align}
\sum_{k^\prime}\left\{ \nabla^2_{kk^\prime} - \delta_{kk^\prime}\frac{l(l+1)}{r^2_k}
\right\}(\bm{W}^p_q)_{k^\prime l} =
-\frac{4\pi(\bm{\rho}^p_q)_{kl}}{r_k\sqrt{w^\textrm{rad}_k}}.
\end{align}
We solve this equation under the boundary condition 
$(\bm{W}^p_q)_{{\color{black}K_\textrm{rad}}l} = \sqrt{w_{\color{black}K_\textrm{rad}}}/R^l$ \cite{McCurdy:2004}, obtaining the mean field as
\begin{eqnarray}
W^p_q(r_k,\theta_j) = \frac{1}{r_k\sqrt{w^\textrm{rad}_k}} \sum_{l=|M^p_q|}^{L_{\rm ee}} G_{j,lM^p_q} ({\color{black} \bm{W}^p_q})_{kl},
\end{eqnarray}
which satisfies $W^p_q(r,\theta, \phi) =
e^{iM^p_q\phi}W^p_q(r,\theta)$.
\\

Finally, the two-electron part $\bm{F}\bm{\varphi}_p$ of Eq.~(\ref{eq:eom_matrix})
is evaluated as
\begin{align}
\label{eq:fock2_act_sphfedvr}
{\color{black}\bm{F}\varphi_p(r_k,\theta_j) = \sum_{oqsr} (D^{-1})_p^o P^{qs}_{or} W^r_s(r_k,\theta_j) \varphi_q(r_k,\theta_j)}
\end{align}
%\end{subequations}
%Finally Eqs.~(\ref{eqs:fock2}), with $\phi$ dependence integrated out, are
%evaluated on the $(r_k,\theta_j)$ grid, 
%\begin{subequations}\label{eq:fock2_sphfedvr}
%\begin{eqnarray}\label{eq:core_fock2_sphfedvr}
%\bm{f}\psi_p(r_k,\theta_j) &=&
%\psi_p(r_k,\theta_j) \sum_{qr}D^r_q W^q_r(r_k,\theta_j) \nonumber \\ &-&
%\frac{1}{2} \sum_{qr} \psi_r(r_k,\theta_j) D^r_q W^q_p(r_k,\theta_j)
%\\
%\label{eq:fock2_act_sphfedvr}
%\bm{F}\varphi_p(r_k,\theta_j) &=& \bm{f}\varphi_p(r_k,\theta_j) \\ &+&
%\sum_{u} \varphi_q(r_k,\theta_j) T^q_p(r_k,\theta_j), \nonumber
%\end{eqnarray}
%\end{subequations}
and transformed back into the spherical-FEDVR basis,
\begin{eqnarray}
\bm{F} \bm{\varphi}_{p,kl} &=& 
\sum_{j=0}^{L_{\rm ee}} G^{-1}_{lm_p,j} \bm{F}\varphi_p(r_k,\theta_j).
\end{eqnarray}
%
%\begin{subequations}\label{eq:fock2_sphfedvr}
%\begin{eqnarray}\label{eq:core_fock2_sphfedvr}
%\bm{f}\psi_p(r_k,\theta_j) &=&
%2\psi_p(r_k,\theta_j)\sum_i W_{ii}(r_k,\theta_j) \nonumber \\
%&-& \sum_i \psi_i(r_k,\theta_j) W_{ip}(r_k,\theta_j),
%\end{eqnarray}
%\begin{eqnarray}\label{eq:fock2_core_sphfedvr}
%\bm{F}^{\rm 2e}_i\varphi_i(r_k,\theta_j) &=& \bm{f}\phi_i(r_k,\theta_j) \\ &+&
%\psi_i(r_k,\theta_j) \sum_{tu}D_{tu} W_{ut}(r_k,\theta_j) \nonumber \\ &-&
%\frac{1}{2} \sum_{tu} \psi_t(r_k,\theta_j) D_{tu} W_{ui}(r_k,\theta_j),\nonumber 
%\end{eqnarray}
%\begin{eqnarray}\label{eq:fock2_act_sphfedvr}
%\bm{F}^{\rm 2e}_t\varphi_t(r_k,\theta_j) &=& \bm{f}\varphi_t(r_k,\theta_j) \\ &+&
%\sum_{uvw} \varphi_u(r_k,\theta_j) W_{vw}(r_k,\theta_j) {\it \Gamma}_{ut,wv}, \nonumber
%\end{eqnarray}
%\end{subequations}
%and transformed back into the spherical-FEDVR basis,
%\begin{eqnarray}
%\bm{F}^{\rm 2e}_p \bm{\varphi}_{p,kl} &=& 
%\sum_{j=0}^{L_{\rm ee}} T^{-1}_{lm_p,j} \bm{F}^{\rm 2e}_p\varphi_p(r_k,\theta_j).
%\end{eqnarray}
%
When we use FC orbitals, {\color{black} their contribution $\bm{F}_{\rm fc}$ to the full operator $\bm{F}$ % to Eq.~(\ref{eq:core_fock2_sphfedvr})
is treated separately,}
\begin{align}\label{eq:core_fock2_sphfedvr_fc}
\bm{F}_{\rm fc}\varphi_p(r_k,\theta_j,t) &= %&\leftarrow&
\varphi_p(r_k,\theta_j,t) V_\textrm{fc}(r_k,\theta_j,t=0) \nonumber \\
&- \sum_i^\textrm{fc} \varphi_i (r_k,\theta_j,t) W^i_p(r_k,\theta_j,t),
\end{align}
where the first Coulomb term {\color{black}$V_\textrm{fc} = 2\sum_i^\textrm{fc} W^i_i$}
is evaluated once in the beginning of the simulation and serves as a
multiplicative operator, while the second exchange term is evaluated at
each time step $t$, with the transformation Eq.~(\ref{eq:fc_vg}) for VG
simulations. The exchange term $W^i_p$ is evaluated only within a sphere $r_k <
R^\textrm{fc}_i$, where $R^\textrm{fc}_i$ is once determined for each FC orbital
$\varphi_i$ so that $|\varphi_i(R^\textrm{fc}_i)|$ is below a given threshold
$\delta^\textrm{fc}$. {\color{black}In this work we use} $\delta^\textrm{fc} = 10^{-15}$.
%Coulomb is calculated only once before the real-time
%simulation (and possibly marged into $\hat{h}^{\rm 1e}$). Poisson's
%equation and grid summation for the frozen-core exchange is evaluated
%within a very small radial box.

%%% I don't understand this part %%% Interestingly, the above discussion gives an insight into how to
%%% I don't understand this part %%% calculate dipole acceleration within the single active electron
%%% I don't understand this part %%% approximation. See attached paper by Gordon, Santra, et al. They argue
%%% I don't understand this part %%% that we should use their Eq. (10) corresponding to our Eq. (A8) and
%%% I don't understand this part %%% Eq. (i) above, rather than their Eq. (6) corresponding to Eq. (ii)
%%% I don't understand this part %%% above. (their Eq. (6) can be transformed into the expectation value of
%%% I don't understand this part %%% the force from the effective potential, i.e., $f_{\rm na}+f_{\rm ca}$).
%%% I don't understand this part %%% Numerical results in this work and the above discussion indicates the
%%% I don't understand this part %%% opposite.

\bibliography{refs.bib}
\end{document}